\documentclass[%
preprint,
superscriptaddress,
preprintnumbers,
nobibnotes,
nofootinbib,
 amsmath,amssymb, 
 aps, prd,
]{revtex4-1}

\usepackage{cancel}
\usepackage{accents}
\usepackage{mciteplus,slashed}
\usepackage{cancel}
\usepackage{dcolumn}
\usepackage{bm}
\usepackage{soul}
\usepackage[caption=false]{subfig}
\usepackage{appendix}
\usepackage{physics,dsfont}
\usepackage{booktabs}
\usepackage{comment}
\usepackage[T1]{fontenc}	
\usepackage{csvsimple}
\usepackage[bookmarksnumbered=true]{hyperref} 
\hypersetup{
    colorlinks = true,
    linkcolor = blue,
    anchorcolor = blue,
    citecolor = blue,
    filecolor = blue,
    urlcolor = blue
    }
\usepackage[section]{placeins}
\usepackage[capitalise]{cleveref}
\usepackage{booktabs}
\usepackage{graphicx}
\usepackage{mathrsfs}
\usepackage{relsize}
\graphicspath{{Figures/}}
\AtBeginDocument{
\heavyrulewidth=.08em
\lightrulewidth=.05em
\cmidrulewidth=.03em
\belowrulesep=.65ex
\belowbottomsep=0pt
\aboverulesep=.4ex
\abovetopsep=0pt
\cmidrulesep=\doublerulesep
\cmidrulekern=.5em
\defaultaddspace=.5em
}
\usepackage[dvipsnames]{xcolor}
\usepackage[normalem]{ulem}
\usepackage{fontawesome} 
\usepackage{bbm}
\usepackage{MnSymbol}
\usepackage{comment}
\usepackage[force]{feynmp-auto}

\def\iu{\mathrm{i}}
\def\e{\mathrm{e}}

\DeclareSymbolFont{cmbrightop}{OT1}{cmbr}{m}{n}
\DeclareMathSymbol{\sfPsi}{\mathalpha}{cmbrightop}{9}

\definecolor{wildstrawberry}{rgb}{1.0, 0.26, 0.64}

\begin{document}

\begin{fmffile}{fmfnotes} 
\title{Vertex corrections and wavefunction renormalization \\ for atoms, nuclei, and other heavy composite particles }

\author{Ryan Plestid}
%\email{rplestid@caltech.edu}
\author{Mark B.\ Wise}
%\email{wise@caltech.edu}
\affiliation{Walter Burke Institute for Theoretical Physics, California Institute of Technology, Pasadena, CA 91125, USA}

\date{\today}

\preprint{CALT-TH/2025-025}

\begin{abstract}
  We study QED corrections to operator matrix elements involving heavy composite particles (e.g., heavy-mesons, nuclei, and atoms). We define a new notion of reducible and irreducible graphs which is useful for systems with many discrete excited states. The equivalence of the LSZ reduction formula and old fashioned perturbation theory is explicitly demonstrated. The self energy and vertex corrections are defined (to all orders), and the one-loop corrections are reduced to operator matrix elements which may be evaluated by hadronic, nuclear, or atomic theorists. The gauge dependence of the various pieces are studied in detail at one loop, and cancellation of spurious contributions are demonstrated in a class of covariant gauges; Coulomb gauge is also discussed. The formalism is applied to superallowed beta decay where the one-loop structure is connected to existing literature based on current algebra techniques. We further identify the well known $O(Z^2\alpha^2)$ isospin breaking correction from the intranuclear Coulomb field as arising from two-loop diagrams. We comment on future applications of our results to the radiative corrections necessary in extractions of $|V_{ud}|$, in particular for corrections that required beyond one-loop order. 
\end{abstract}

\maketitle 

%\tableofcontents

\vfill
\pagebreak

\section{Introduction \label{Intro} }
There is a growing interest in high precision measurements of processes involving atoms, nuclei, and heavy hadrons \cite{Afanasev:2023gev,Hardy:2020qwl,Zwicky:2021olr,Tomalak:2022xup,Pastore:2009is,Cornella:2022ubo,PREX:2021umo,Hill:2023acw,Seng:2022cnq,Cirigliano:2024rfk,Plestid:2024jqm,Beneke:2017vpq,Tomalak:2020zfh,Combes:2024pvm,Cirigliano:2023fnz,Seng:2023cgl,Baroni:2021vrc,Hill:2016bjv,Pastore:2008ui,VanderGriend:2025mdc,Seng:2024zuc,Lovato:2020kba,Plestid:2024xzh,Cirigliano:2018yza,Richardson:2023vyf,Gennari:2024sbn,Epelbaum:2019kcf,Seng:2022cnq,Cirigliano:2024rfk,Cirigliano:2022oqy,Tomalak:2021hec,Cirigliano:2024nfi,Hill:2023bfh,Rocco:2020jlx,MUonE:2016hru,Pastore:2011ip,Hill:2010yb,Borah:2024ghn,CREX:2022kgg,Ruso:2022qes,Richardson:2025aqy}. For precision targets of $\sim 1\%$ or better, an understanding of quantum electrodynamic (QED) corrections at least through one-loop order is required. Analysis of these systems are complicated by the complexity of the bound-state dynamics, despite their underlying constituents being relatively well understood. 

As a concrete example, in leptonic reactions involving nuclei, existing analyses sometimes restrict themselves to leptonic corrections and so-called ``box diagrams'' in which a photon is exchanged between a charged lepton and the hadronic system. For neutral current reactions, these topologies are separately gauge invariant, as are the nuclear vertex corrections. Nevertheless a complete analysis through $O(\alpha)$ (with $\alpha$ the fine structure constant) demands the inclusion of the corrections to the nuclear vertex, and it is desirable to have a formalism that both properly defines this object, and facilitates realistic calculations given nuclear wavefunctions for the initial and final state.\!\footnote{See Ref.~\cite{Preparata:1968ioj,Sirlin:1977sv,Seng:2021syx} for a discussion of some of these issues in the language of current algebra. \label{footnote-1}} 

For charged current reactions, such as neutrino nucleus scattering or beta decay, the problem of the nuclear vertex correction is more pressing. Because the charge of the initial and final nucleus differ, the leptonic, box, and nuclear vertex corrections will all in general carry gauge dependence. Therefore, in order to ensure gauge invariance of the amplitude, the gauge dependencies of the various pieces must be properly understood. 

In this paper we study the construction of the Lehman-Symanzik-Zimmerman (LSZ) reduction formula \cite{Lehmann:1954rq} for bound states composed of electrically charged constituents. The resulting perturbation theory can be expressed entirely in terms of nuclear matrix elements taken between the initial state $\ket{A}$ and the final state $\ket{B}$ defined in the absence of electromagnetism. The results can also be derived using time-independent perturbation theory. Next the gauge dependence of the vertex correction is studied in covariant gauges, and a comparison is made with a calculation in Coulomb gauge. We then apply our results to the theory of beta decay. We match onto existing results at one-loop, and identify the $Z^2\alpha^2$ isospin breaking correction from the overlap of nuclear wavefunctions as a subset of two-loop contributions. We finally summarize our results and present an outlook for future work.

\pagebreak
\vfill 

\section{Structure of the LSZ reduction formula}
Let us review the standard LSZ construction \cite{Lehmann:1954rq,Weinberg:1995mt}. Consider the matrix element of an arbitrary operator, $\mathcal{O}$, between single particle states $A$ and $B$. We denote exact eigenstates of the full theory\footnote{Our notation is taken from  Refs.~\cite{Miller:2008my,Miller:2009cg}, but follows an opposite convention where $|A)$ is the fully interacting state. We find this choice convenient since we work in perturbation theory and mostly use $\ket{A}$.} by $|A)$ and $|B)$, 
\begin{equation}
    \mathcal{M}= (B(p')|\mathcal{O}|A(p))~. 
\end{equation}
Let us suppose we have interpolating operators $\psi_A^\dagger$ and $\psi_B^\dagger$ which satisfy 
\begin{align}
    (A(p)|\psi_A^\dagger|\Omega) &= \sqrt{Z_A} \xi_A^\dagger(p)~, ~\\
    (B(p)|\psi_B^\dagger|\Omega) &= \sqrt{Z_B} \xi_B^\dagger(p)~,
\end{align}
where $|\Omega)$ is the full interacting vacuum, $Z_{A,B}$ are real numbers, and $\xi_{A,B}$ are external wavefunctions which transform appropriately under the Lorentz group (e.g., Dirac spinor wavefunctions for spin-$1/2$ particles). Operators at a position $x$ are defined by acting with the generator of translations $P$, which gives,  $\psi_{A,B}(x)=\e^{\iu P\cdot x} \psi_{A,B} \e^{-\iu P\cdot x}$.

The LSZ reduction formula then tells us that 
\begin{equation}
  \begin{split}
     \mathcal{M} = \lim_{p_0'\rightarrow E_B} \lim_{p_0\rightarrow E_A} \int &\dd^4x \int \dd^4 y ~~ \e^{\iu p'\cdot x} \e^{-\iu p\cdot y}  \\
      &\times Z_B^{-1/2} \xi_B^\dagger(p')  \overset{\rightarrow}{\mathcal{D}_x}(\Omega|T\{ \psi_B(x) \psi_A^\dagger(y) \mathcal{O}(0)\}|\Omega)\overset{\leftarrow}{\mathcal{D}_y} \xi_A(p) Z_A^{-1/2}~,
    \end{split}
\end{equation}
where $\mathcal{D}_x$ is the appropriate differential operator for particle $B$ (as dictated by its spin) and likewise for $\mathcal{D}_y$ and particle $A$.  Accounting for factors of $Z_A$ and $Z_B$ from the residues on the external legs, this may be conveniently summarized by 
\begin{equation}
    \label{LSZ-heutristic}
     \mathcal{M} = \sqrt{Z_A} \sqrt{Z_B} \times \qty[\xi_B^\dagger(p')~ G_{\rm amp.}(p',p) ~\xi_A(p)]~,
\end{equation}
where $G_{\rm amp}(p',p)$ is the amputated Greens function evaluated at on-shell kinematics for a particle $A$ transitioning to another particle $B$ as mediated by the operator $\mathcal{O}$.

\vfill
\pagebreak 

\subsection{Perturbative expansion}
For conceptual simplicity let us consider nuclei, which are bound in the absence of electromagnetism. In the limit that the electromagnetic coupling constant tends to zero, $e\rightarrow 0$, the system is described by the strong interaction $H_S$. Neglecting the  target recoil kinetic energy (we will always assume the bound state is heavy enough that this is a good approximation) the full Hamiltonian is then
\begin{equation}
    H_{\rm full}= H_S + H_{\gamma}+H_{\rm int}~, 
\end{equation}
where $H_{\rm \gamma}$ is the free photon Hamiltonian, and $H_{\rm int}$ contains all interactions between the constituent nucleons and photons i.e., the current operators. Next, define a modified interaction picture, where $H_{\rm int}$ is treated as the interaction Hamiltonian, and $H_S$ is treated as the zeroth order Hamiltonian. Let us illustrate how this picture works with an example in the limit of $e\rightarrow 0$. Consider the operator, $\psi$, that annihilates the vacuum, $\psi\ket{0}=0$. The two point function of this operator is given by 
\begin{equation}
    \int \dd^4x~\e^{\iu p\cdot x} \mel{0}{T\{ \psi(x) \psi^\dagger(0)\}}{0} =  \int \frac{\dd^3 k}{(2\pi)^3} \mel{0}{\psi_{\vb{p}}\frac{\iu}{p_0 - H_S+\iu 0} \psi^\dagger_{\vb{k}}}{0} ~, 
\end{equation}
where only one time ordering contributes because $\psi\ket{0}=0$. We define 
\begin{align}
    \psi^\dagger_{\vb{k}} &= \int \dd^3x ~\e^{\iu \vb{k} \cdot \vb{x}} \psi^\dagger(t=0,\vb{x})~,\\
    \psi_{\vb{k}} &= \int \dd^3x ~\e^{-\iu \vb{k} \cdot \vb{x}} \psi^\dagger(t=0,\vb{x})~
\end{align}
such that $\psi^\dagger_{\vb{k}}$ creates,  and $\psi_{\vb{k}}$ annihilates, states with momentum $\vb{k}$.

The wavefunction renormalization $Z_{A,B}$, and the amputated Greens function can be computed order by order in the electromagnetic coupling constant $e$ using,
\begin{equation}
    \label{Gell-Mann-Low}
    \begin{split}
      (\Omega|T\{ \psi_B(y) \psi_A^\dagger(x) \mathcal{O}(0)\} |\Omega)= \frac{\mel{0}{T\{ \e^{\iu \int \dd^4 z \mathcal{L}_{\rm int} }  \psi_B(y) \psi_A^\dagger(x) \mathcal{O}(0)\} }{0}}{\mel{0}{T\{\e^{\iu \int \dd^4 z \mathcal{L}_{\rm int} }\}}{0} }~.
    \end{split}
\end{equation}
The left-hand side involves the fully interacting vacuum $|\Omega)$ and operators in the Heisenberg picture (with respect to $H_{\rm full}$), while the right-hand side involves operators in the interaction picture defined above, and the corresponding vacuum $\ket{0}$ defined by $(H_S+H_\gamma) \ket{0}=0$. 

We always consider bound states with at least one heavy constituent so that $\psi_A\ket{0}=\psi_B\ket{0}=0$. This identity is related to the conservation of particle number for non-relativistic particles, and applies even when light degrees of freedom (e.g., pions in a nucleus or light quarks in a $\overline{B}$ meson) are present.  

\subsection{Construction of the interpolating field}
When considering non-relativistic bound-states (such as nuclei) a natural interpolating operator is the solution of the $N$-body Schr\"odinger equation in the $e\rightarrow 0$ limit
\begin{equation}
    \psi_A^\dagger = \int \frac{\dd^3 k}{(2\pi)^3} \int \qty{ \frac{\dd^3p_1}{(2\pi)^3} \ldots \frac{\dd^3p_N}{(2\pi)^3} } \Psi\qty(\vb{p}_1,\ldots,\vb{p}_N)  a^\dagger_{\vb{p}_1} \ldots a^\dagger_{\vb{p}_N} (2\pi)^3\delta^{(3)}(\Sigma \vb{p}_i - \vb{k})~,
\end{equation}
where $\Psi\qty(\vb{p}_1,\ldots,\vb{p}_N)$ is the many-body wavefunction and the center of mass momentum $\vb{k}$ has been made explicit. We take our states to be normalized as $\braket{A(\vb{p}')}{A(\vb{p})}=(2\pi)^3 \delta^{(3)}(\vb{p}-\vb{p}')$. This operator has the property that $\mel{B(p)}{\psi_A^\dagger}{0} = \delta_{AB} \xi_A^\dagger(p)$  i.e., there is perfect overlap with the state $A$ and total orthogonality to all other states $B\neq A$. 

In a more general setting (e.g., with dynamical pions) it is unlikely that one can define such a ``perfect'' interpolating operator constructively.  Nevertheless, such an object exists formally, as can be easily seen (up to a trivial re-scaling) by considering a general interpolating field and acting with the projector, $\int \dd\vb{p} \ketbra{A(\vb{p})}{A(\vb{p})}$ (with $A$ {\it only} the ground state). Therefore, in what follows we use an interpolating field that satisfies $\mel{B(p)}{\psi_A^\dagger}{0} = \delta_{AB} \xi_A^\dagger(p)$. 

\subsection{Irreducible graphs and Feynman rules}
In standard treatments of the LSZ reduction formula the notion of one-particle irreducible (1PI) graphs is very important. Naively infinite results (i.e., higher order poles) for the two-point function are resummed and known to be related to mass and wavefunction renormalization which are both expressed in terms of the 1PI self-energy. 

In the present context, 1PI is hard to define. The classification of single-particle states will, in general, not be stable under electromagnetic perturbations; a bound state near threshold may move into the continuum or vice-versa. Moreover, the number of single-particle excited states in the spectrum may be very large (or infinite for atoms) rendering the dimensionality of a 1PI self-energy matrix intractable; implementing the 1PI scheme would be almost as hard as solving the full many-body problem. It is therefore desirable to develop a modified notion of 1PI graphs for applications to bound states. 

Let us consider the two-point function, and assume a non-degenerate spectrum; for simplicity take $A$ and $B$ to have spin-$0$. Due to the non-relativistic nature of (at least one of) the constituents, and the orthogonality relation $\mel{B(p)}{\psi_A^\dagger}{0} = \delta_{AB}$, all non-zero graphs begin and end with the state $A$. Any graph with an $A$-propagator will yield a higher order pole in the vicinity of $p_0 \rightarrow E_A^{(0)}$ where $H_S \ket{A} = E_A^{(0)}$ defines the unperturbed energy of $A$. Iterating gives, 
\begin{equation}
    \begin{split}
    G(p)
    = \frac{\iu}{p_0-E_A^{(0)}}+  \frac{\iu}{p_0-E_A^{(0)}}&(-\iu \Sigma_2) \frac{\iu}{p_0-E_A^{(0)}} +\ldots~,
    \end{split}
\end{equation}
which (following the standard argument) can be resummed,
\begin{equation}
    G(p) = \frac{\iu}{p_0-H_S-\Sigma_2(p_0)+\iu 0} ~. 
\end{equation}
By the above analysis, the self energy, $\Sigma_2(p_0)$, does not contain any isolated ground state propagators (i.e., outside of a photon loop). This motivates the introduction of the projectors
\begin{align}
    \mathbb{P}_{A} &=\int \frac{\dd^3 p}{(2\pi)^3}\ketbra{A(\vb{p})}{A(\vb{p})}~,\\
    \mathbb{P}_{\slashed{A}} &= 1- \int \frac{\dd^3 p}{(2\pi)^3}\ketbra{A(\vb{p})}{A(\vb{p})}~,
\end{align}
which satisfy $\mathbb{P}_{A}+ \mathbb{P}_{\slashed{A}}=1$. 

Using this projector we can define the notion of {\it external particle irreducible} (EPI) graphs. For a process with an external single-particle state $A$, a graph is EPI if it cannot be reduced into two separate pieces by cutting an $A$-propagator in the graph. The self energy is defined by all EPI graphs, which differs from the conventional definition in terms of 1PI graphs. Using different projectors in place of $\mathbb{P}_{\slashed{A}}$ other irreducible graph schemes can be constructed (for example 1PI corresponds to a projector that removes all single-particle states) in direct analogy with degenerate perturbation theory in non-relativistic quantum mechanics (see \cref{OFPT}). 

The notion of EPI naturally motivates the following Feynman rules. There is a standard photon propagator (which we write momentum $q$) and three hadronic propagators\footnote{The double line and dashed double line include both discrete and continuum excited states. The latter may be thought of as accounting for loops of strongly interacting particles.} (which we write with momentum $p$) 
\begin{align}
    \raisebox{-3.33pt}{\begin{fmfgraph*}(10,3)
    \fmfleft{i1}
    \fmfright{o1}
    \fmf{photon}{i1,o1}
    \end{fmfgraph*}}
    &
    ~~=~~D_{\mu\nu}(q) ~, \\[3pt]
    \raisebox{-3.33pt}{\begin{fmfgraph*}(10,3)
    \fmfleft{i1}
    \fmfright{o1}
    \fmf{double}{i1,o1}
    \end{fmfgraph*}}
    &
    ~~=~~\frac{\iu}{p_0-H_S+\iu 0} ~, \\[3pt]
    \raisebox{-3.33pt}{\begin{fmfgraph*}(10,3)
    \fmfleft{i1}
    \fmfright{o1}
    \fmf{dbl_dashes}{i1,o1}
    \end{fmfgraph*}}  &~~=~~\frac{\iu}{p_0-H_S+\iu 0}\mathbb{P}_{\slashed{A}}~, \\[3pt]
    \raisebox{-3.33pt}{\begin{fmfgraph*}(10,3)
    \fmfleft{i1}
    \fmfright{o1}
    \fmf{plain}{i1,o1}
    \end{fmfgraph*}}  &~~=~~\frac{\iu}{p_0-H_S+\iu 0}\mathbb{P}_{A}~.
\end{align}
These propagators are redundant, since~~
$
    \raisebox{-3.33pt}{\begin{fmfgraph*}(10,3)
    \fmfleft{i1}
    \fmfright{o1}
    \fmf{plain}{i1,o1}
    \end{fmfgraph*}}~+~\raisebox{-3.33pt}{\begin{fmfgraph*}(10,3)
    \fmfleft{i1}
    \fmfright{o1}
    \fmf{dbl_dashes}{i1,o1}
    \end{fmfgraph*}} ~=~ \raisebox{-3.33pt}{\begin{fmfgraph*}(10,3)
    \fmfleft{i1}
    \fmfright{o1}
    \fmf{double}{i1,o1}
    \end{fmfgraph*}}~,
$
and in practice only the dashed and solid double line propagators are needed for EPI graphs. 

Next, define interaction vertices for photons with momentum $q_\mu =(\omega,\vb{q})$. These are operator valued functions of $\vb{q}$ (but not $\omega)$. For example the current and seagull operators for non-relativistic minimally coupled particles are drawn as, 
\begin{align}
    \raisebox{-6pt}{
    \begin{fmfgraph*}(6,6)
        \fmfbottom{b1}
        \fmftop{t1}
        \fmf{photon}{t1,b1}
        \fmfv{decoration.shape=circle,decoration.filled=empty}{b1}
    \end{fmfgraph*}
    }~~&= (\iu e) J_\mu(\vb{q})~,\\[9pt]
    \raisebox{-6pt}{
    \begin{fmfgraph*}(6,6)
        \fmfbottom{b1}
        \fmftop{t1,t2}
        \fmf{photon}{t1,b1}
        \fmf{photon}{t2,b1}
        \fmfv{decoration.shape=circle,decoration.filled=empty}{b1}
    \end{fmfgraph*}
    }~~&= (\iu e^2) S_{\mu\nu}(\vb{q}_1,\vb{q}_2)~.
\end{align}
More generally, interactions with photons can involve the field strength $F_{\mu\nu}$ coupled to some hadronic operator defined at $t=0$ e.g., $T_{\mu\nu}(\vb{q}_1,\vb{q}_2,\vb{q}_3)$. If derivatives act on the photon field the Feynman rule includes factors of $q_\mu$ (including dependence on $\omega$) but all hadronic operators are evaluated at fixed three-momenta. 

External lines denote either the initial, $\ket{A}$, or final, $\bra{B}$, state
\begin{align}
    \raisebox{-3pt}{\begin{fmfgraph*}(10,3)
        \fmfleft{i1}
        \fmfright{o1}
        \fmfv{label=$A$}{i1}
        \fmf{plain}{i1,o1}
        \fmfv{decoration.shape=circle,decoration.filled=hatched,d.si=0.5w}{o1} 
    \end{fmfgraph*}
    }~~~~~~
    &= ~~~~\ket{A} ~,\\[8pt]
    \raisebox{-3pt}{\begin{fmfgraph*}(10,3)
        \fmfleft{i1}
        \fmfright{o1}
         \fmfv{label=$B$}{o1}
        \fmf{plain}{i1,o1}
        \fmfv{decoration.shape=circle,decoration.filled=hatched,d.si=0.5w}{i1} 
    \end{fmfgraph*}
    }~~~~~~~~
    &= ~~~~\bra{B}~,
\end{align}
where the hatched blob represents the rest of the graph. External polarization spinors, $\xi_{A,B}(p)$, are not needed since these are packaged into the states $\ket{A}$ and $\bra{B}$. 

\subsection{Equivalence with time-independent perturbation theory \label{OFPT} }
The above discussion can be directly related to time-independent perturbation theory. Consider the matrix element $(B| \mathcal{O}(\vb{k})|A)$. As above, exact eigenstates are denoted by $|A)$ and $|B)$ and the unperturbed states by $\ket{A}$ and $\ket{B}$. Let us solve for $|A)$ as a sum over states $\ket{A_n}$, $\ket{A_n}\ket{1\gamma}$,  $\ket{A_n}\ket{2\gamma}$ etc., and then compute $(B|\mathcal{O}(\vb{k})|A)$. 

Since wavefunction renormalization is naturally defined around the exact energy eigenstates, it is convenient to use Brillouin-Wigner \cite{Littlejohn:2019nqs} (rather than Rayleigh-Schr\"odinger) perturbation theory. Let us write the Hamiltonian for the system as $H=H_0 + H_{\rm int}$ where 
\begin{equation}
    H_0 = H_S+H_\gamma~,
\end{equation}
such that $H_\gamma \ket{\gamma(\vb{q})} = |\vb{q}|\ket{\gamma(\vb{q})}$. The state $|A)$ satisfies the Lippmann-Schwinger-like equation \cite{Littlejohn:2019nqs},
\begin{equation}
    |A) = \sqrt{Z_A} \ket{A}  + \frac{1}{E_A - H_0} \mathbb{P}_{\slashed{A}}^{(\gamma)} H_{\rm int} |A)~,
\end{equation}
where $\mathbb{P}_{\slashed{A}}^{(\gamma)} = 1-\int \dd \vb{p} \ketbra{A(\vb{p});0\gamma}{A(\vb{p});0\gamma}$ and $E_A$ is the exact energy of $|A)$. The matrix element is then given by 
\begin{align}
    (B|\mathcal{O}(\vb{k})|A) &= \sqrt{Z_A}\sqrt{Z_B} \mel{B}{\mathcal{O}(\vb{k})}{A}  
    + 
    (B|H_{\rm int}\frac{1}{E_B - H_0} \mathbb{P}_{\slashed{B}}^{(\gamma)} \mathcal{O}(\vb{k})\mathbb{P}_{\slashed{A}}^{(\gamma)}\frac{1}{E_A - H_0}  H_{\rm int} |A)\\
    &+ \sqrt{Z_A}(B|H_{\rm int}\frac{1}{E_B - H_0} \mathbb{P}_{\slashed{B}}^{(\gamma)}\mathcal{O}(\vb{k})\ket{A} 
  + \sqrt{Z_B}
    \bra{B} \mathcal{O}(\vb{k})\mathbb{P}_{\slashed{A}}^{(\gamma)}\frac{1}{E_A - H_0} H_{\rm int} |A)~.\nonumber
\end{align}
The wavefunction renormalization is fixed by demanding the state have unit norm $(A|A)=1$,
\begin{equation}
    Z_A= 1-(A|H_{\rm int} \mathbb{P}_{\slashed{A}}^{(\gamma)} \frac{1}{E_A - H_0} \frac{1}{E_A - H_0} \mathbb{P}_{\slashed{A}}^{(\gamma)} H_{\rm int} |A)~.
\end{equation}
In this section it is important to recall that $\mathbb{P}_{\slashed{A}}^{(\gamma)} \ket{A}\ket{n\gamma}= \ket{A}\ket{n\gamma}$ for $n\geq 1$. At first order in perturbation theory, since $H_{\rm int}$ creates a single photon, we may replace $E_A-H_0 \rightarrow -|\vb{q}| -(H_S-E_A)$ and $E_B-H_0 \rightarrow -|\vb{q}| -(H_S-E_B)$ in the above expressions by acting with $H_\gamma$ on the photon states that appear after acting with $H_{\rm int}$. 

It can be readily checked, after performing the integral over $\dd \omega$ and picking up the pole from the photon propagator, that these results agree with those derived using the LSZ reduction formula and the EPI scheme defined above. Other schemes, in which a set of discrete states are counted as ``external'' are discussed in \cref{app:deg-PT}. Moreover this presents a direct and immediate connection to existing formulations of isospin breaking in nuclear beta decays where only the Coulomb interaction was retained \cite{Miller:2008my,Miller:2009cg}. 

\section{Fixed order results}
It proves convenient to introduce the notation, 
\begin{equation}
    \mel{B}{O_1(\vb{k}_1) \ldots O_N(\vb{k}_N)}{A} \equiv  \frac{\mel{B(\vb{p}_B)}{O_1(\vb{k}_1) \ldots O_N(\vb{k}_N)}{A(\vb{p}_A)}}{\braket{B(\vb{p}_B+\vb{K})}{A(\vb{p}_A)}} ~,
\end{equation}
where $\vb{K}=\sum_i \vb{k}_i$ such that $\mel{B}{\ldots}{A}$ is always a smooth function of $\vb{k}_1\ldots \vb{k}_N$. Therefore, whenever matrix elements without explicit momenta appear below, the above definition is implied. This corresponds to the $\langle \ldots \rangle$ notation in Ref.~\cite{Plestid:2025ojt}, but it is convenient to retain $A$ and $B$ explicitly in the notation when considering reactions involving the weak charged current where the initial and final nuclei differ. 
\subsection{Wavefunction renormalization}
We are now in position to compute the wavefunction renormalization. Analysis of the resummed propagator gives, 
\begin{equation}
    Z_A = \qty[\frac{1}{1-\Sigma_2'}]_{p_0=E_A}~, 
\end{equation}
which holds at all orders in perturbation theory. Let us compute the one-loop approximation. Begin with the one-loop self-energy,
\begin{align}
        -\iu \Sigma_2^{(1)}(p_0) 
        &= \begin{fmfgraph*}(20,10)
        \fmfleft{i1,i2}
        \fmfright{o1,o2} 
        \fmf{phantom}{i1,o2}
        \fmf{plain}{i1,v1}
        \fmf{double}{v1,v3,v4,v2}
        \fmf{plain}{v2,o1}
        \fmfv{decoration.shape=circle,decoration.filled=empty}{v1}
        \fmfv{decoration.shape=circle,decoration.filled=empty}{v2}
        \fmffreeze
        \fmf{photon,left=1.2}{v1,v2}
        %\fmf{photon,left=2}{v1,v2}
    \end{fmfgraph*} 
    + 
    \begin{fmfgraph*}(15,6)
        \fmfleft{i1,i2}
        \fmfright{o1,o2}
        \fmf{phantom}{i2,p1,p2,p3,o2}
        \fmf{plain}{i1,v1,o1}
        \fmfv{decoration.shape=circle,decoration.filled=empty}{v1}
        \fmffreeze
        \fmf{photon,left=0.3}{v1,p1}
        \fmf{photon,right=0.3}{v1,p3}
        \fmf{photon,left=0.3}{p1,p3}
        %\fmf{photon,left=2}{v1,v2}
    \end{fmfgraph*} \\
    &=(-\iu e^2) \int \frac{\dd^4q}{(2\pi)^4}
    D_{\mu\nu}(q)  \qty[\mel{A}{J_\nu(-\vb{q}) \frac{1}{p_0+\omega-H}J_\mu(\vb{q})}{A}  + \frac12 \mel{A}{S_{\mu\nu}(\vb{q},-\vb{q})}{A} ]~.\nonumber
\end{align}
where we use $\omega=q_0$ throughout the rest of the paper. 
The factor of $1/2$ for the seagull vertex is related to a symmetry factor. 
Next, differentiate with respect to $p_0$ and set $p_0=E_A$, 
\begin{equation}
    \begin{split}
      \qty[ \dv{p_0} \Sigma_2^{(1)}]_{p_0=E_A} &=  \begin{fmfgraph*}(20,10)
        \fmfleft{i1,i2}
        \fmfright{o1,o2} 
        \fmf{phantom}{i1,o2}
        \fmf{plain}{i1,v1}
        \fmf{double}{v1,v3,v5,v4,v2}
        \fmf{plain}{v2,o1}
        \fmfdot{v5} \fmfv{decoration.shape=circle,decoration.filled=empty}{v1}
        \fmfv{decoration.shape=circle,decoration.filled=empty}{v2}
        \fmffreeze
        \fmf{photon,left=1.2}{v1,v2}
        %\fmf{photon,left=2}{v1,v2}
    \end{fmfgraph*}\\
        &=(-e^2)\times  \int \frac{\dd \omega}{(2\pi)} D_{\mu\nu}(q)  \mel{A}{J_\nu(-\vb{q}) \qty(\frac{1}{\omega-H_A+\iu 0})^2 J_\mu(\vb{q})}{A}~,
    \end{split}
\end{equation}
where the black dot, $\bullet=(-1)\times \mathds{1}$, is proportional to the identity operator. Next, consider the diagrammatic expansion at second order. There are the familiar topologies from scalar-QED (shown in the first line) and then an additional interaction of the lower-order self energy with the dashed propagator (shown in the second line), 
\begin{equation}
    \begin{split}
   -\iu \Sigma_2^{(2)}=
   &\!\begin{fmfgraph*}(30,10)
        \fmfleft{i1,i2}
        \fmfright{o1,o2} 
        \fmf{phantom}{i1,o2}
        \fmf{plain}{i1,h0,v1}
        \fmf{double}{v1,h1,h2,h3,v2}
        \fmf{double}{v2,h7,h8,h9,v3}
        \fmf{double}{v3,h4,h5,h6,v4}
        \fmf{plain}{v4,h10,o1} \fmfv{decoration.shape=circle,decoration.filled=empty}{v1}
        \fmfv{decoration.shape=circle,decoration.filled=empty}{v2}
        \fmfv{decoration.shape=circle,decoration.filled=empty}{v3}
        \fmfv{decoration.shape=circle,decoration.filled=empty}{v4}
        \fmffreeze
        \fmf{photon,left=0.75}{v1,v3}
        \fmf{photon,left=0.75}{v2,v4}%\fmf{photon,left=2}{v1,v2}
    \end{fmfgraph*}
    \!+\!
    \begin{fmfgraph*}(30,10)
        \fmfleft{i1,i2}
        \fmfright{o1,o2} 
        \fmf{phantom}{i1,o2}
        \fmf{plain}{i1,h0,v1}
        \fmf{double}{v1,h1,h2,h3,v2}
        \fmf{double}{v2,h7,h8,h9,v3}
        \fmf{double}{v3,h4,h5,h6,v4}
        \fmf{plain}{v4,h10,o1} 
        \fmfv{decoration.shape=circle,decoration.filled=empty}{v1}
        \fmfv{decoration.shape=circle,decoration.filled=empty}{v2}
        \fmfv{decoration.shape=circle,decoration.filled=empty}{v3}
        \fmfv{decoration.shape=circle,decoration.filled=empty}{v4}
        \fmffreeze
        \fmf{photon,left=0.5}{v1,v4}
        \fmf{photon,left=1}{v2,v3}%\fmf{photon,left=2}{v1,v2}
    \end{fmfgraph*}
    + ~~\ldots ~~\\[12pt]
    &~~~~~~\!+\!
    \begin{fmfgraph*}(30,5)
        \fmfleft{i1,i2}
        \fmfright{o1,o2} 
        \fmf{phantom}{i1,o2}
        \fmf{plain}{i1,v1}
        \fmf{dbl_dashes}{v1,h8,v2}
        \fmf{plain}{v2,o1} 
        \fmf{phantom}{i2,p1,p2,p3,p4,p5,p6,p7,o2}
        \fmfv{decoration.shape=circle,decoration.filled=empty}{v1,v2}
        \fmffreeze
        \fmf{photon,left=0.3}{v1,p1}
        \fmf{photon,right=0.3}{v1,p3}
        \fmf{photon,left=0.3}{p1,p3}
        \fmf{photon,right=0.3}{v2,p7}
        \fmf{photon,left=0.3}{v2,p5}
        \fmf{photon,right=0.3}{p7,p5}
    \end{fmfgraph*}
    \!+\!
    ~~~\ldots ,
    \end{split}
\end{equation}
with ellipses denoting other topologies. Notice the graphs involving iterations of $-\iu \Sigma^{(1)}_2(p_0)$ are EPI but not 1PI.  

When considering gauge-dependence, it is useful to introduce the operators $\mathscr{J}^{(\pm)}_\mu(q)$, defined by
\begin{equation}
    \label{curly-J-def}
    \mathscr{J}_\mu^{(\pm)}(q) = J_{\perp\mu}(\vb{q}) \pm \frac{1}{\pm\omega+\iu 0}(\vb{q}\cdot \vb{J})  v_\mu~.
\end{equation}
where $v_\mu$ is the four-velocity of the nuclear target (i.e., $v_\mu = (1,0,0,0)$ in the rest frame) and $\omega=v\cdot q$. These operators satisfy $q^\mu \mathscr{J}_\mu^{(\pm)}(q)=0$, and arise naturally from the identities \cite{Plestid:2025ojt}, 
\begin{equation}
    \label{trick-1}
    \begin{split}
     &\frac{1}{\omega-H_A+\iu 0}J_\mu(\vb{q}) \ket{A}=\qty[\frac{1}{\omega +\iu 0}\rho(\vb{q})v_\mu +  \frac{1}{\omega-H_A +\iu 0}\mathscr{J}_{\perp\mu}^{(+)}(q)]\ket{A}  ~,
    \end{split}
\end{equation}
\begin{equation}
    \label{trick-2}
    \begin{split}
     &\bra{B}J_\mu(\vb{q})\frac{1}{-\omega-H_B +\iu 0}=\bra{B}\qty[\rho(\vb{q})v_\mu\frac{1}{-\omega +\iu 0} +  \mathscr{J}_\mu^{(-)}(q)\frac{1}{-\omega-H_B +\iu 0}]  ~,
    \end{split}
\end{equation}
where $H_{A,B}=H_S-E_{A,B}^{(0)}$ with $H_S\ket{A,B}=E_{A,B}^{(0)}\ket{A,B}$. For example, when evaluated on-shell, the one-loop self-energy can be symmetrized under $q\leftrightarrow -q$ in which case it can be written in precisely the same form as the Cottingham formula \cite{Cottingham:1963zz}, 
\begin{equation}
    -\iu \Sigma^{(1)}_2(p_0=E_A) = \frac{(-e^2)}{2} \int \frac{\dd^4 q}{(2\pi)^4} D_{\mu\nu}(q) H_{\mu\nu}(q,-q)~,
\end{equation}
where $H_{\mu\nu}$ is the doubly-virtual Compton tensor. A manifestly gauge invariant decomposition of this object using \cref{trick-1,trick-2} is given in Ref.~\cite{Plestid:2025ojt}. 
\subsection{Vertex correction}
Next, consider the amputated Greens function $G_{\rm amp}(p',p)$. At tree level, only the matrix element of $\mathcal{O}(\vb{k})$ is required,
\begin{equation}
   \xi^\dagger_B(p') G_{\rm amp}^{(0)}(p',p)\xi_A(p) = \mel{B}{\mathcal{O}(\vb{k})}{A} = 
    ~~~
    \raisebox{-4pt}{\begin{fmfgraph*}(9,3)
        \fmfleft{i1,i2}
        \fmfright{o1,o2}
        \fmftop{t1}
        \fmf{plain}{i1,t1,o1}
        \fmfv{label=$A$}{i1}
        \fmfv{label=$B$}{o1}
        \fmfv{d.sh=square,d.si=0.2w}{t1}
    \end{fmfgraph*}
    }
\end{equation}
Next, at one-loop order there are five graphs, 
\begin{equation}
    \begin{split}
   \xi^\dagger_B(p')  G_{\rm amp}^{(1)}(p',p)\xi_A(p)
   &= \mathcal{A}^{(1)}_a+ \mathcal{A}^{(1)}_b + \mathcal{A}^{(1)}_c + \mathcal{A}^{(1)}_d + \mathcal{A}^{(1)}_e~, 
    \end{split}
\end{equation}
which are given diagrammatically by 
\begin{align}
    \mathcal{A}^{(1)}_a ~&=~~~~
    \raisebox{-4pt}{\begin{fmfgraph*}(18,6)
        \fmfleft{i1,i2}
        \fmfright{o1,o2}
        \fmftop{t1}
        \fmf{plain}{i1,v1}
        \fmf{double}{v1,h1,t1,h2,v2}
        \fmf{plain}{v2,o1}
        \fmfv{label=$A$}{i1}
        \fmfv{label=$B$}{o1}
        \fmfv{d.sh=square,d.si=0.1w}{t1}
        \fmffreeze
        \fmf{photon}{v1,v2}
        \fmfv{decoration.shape=circle,decoration.filled=empty,d.si=0.1w}{v1}
        \fmfv{decoration.shape=circle,decoration.filled=empty,d.si=0.1w}{v2}
    \end{fmfgraph*}
    }~~~,\\[18pt]
    \label{amp-b-c}
    \mathcal{A}^{(1)}_b + \mathcal{A}^{(1)}_c~&=~~~~
    \raisebox{-8pt}{\begin{fmfgraph*}(18,9)
        \fmfleft{i1,i2}
        \fmfright{o1,o2}
        \fmftop{t1}
        \fmf{plain}{i1,v1}
        \fmf{double}{v1,h1,v2}
        \fmf{dbl_dashes}{v2,h2,h3,t1}
        \fmf{plain}{t1,o1}
        \fmfv{label=$A$}{i1}
        \fmfv{label=$B$}{o1}
        \fmfv{d.sh=square,d.si=0.1w}{t1}
        \fmffreeze
        \fmf{photon,left=1.5}{v1,v2}
        \fmfv{decoration.shape=circle,decoration.filled=empty,d.si=0.1w}{v1}
        \fmfv{decoration.shape=circle,decoration.filled=empty,d.si=0.1w}{v2}
    \end{fmfgraph*}
    }~~~~+~~~~
    \raisebox{-8pt}{\begin{fmfgraph*}(18,9)
        \fmfleft{i1,i2}
        \fmfright{o1,o2}
        \fmftop{t1}
        \fmf{plain}{i1,v1}
        \fmf{phantom,tension=100}{v1,h1,v2}
        \fmf{dbl_dashes}{v2,t1}
        \fmf{plain}{t1,o1}
        \fmfv{label=$A$}{i1}
        \fmfv{label=$B$}{o1}
        \fmfv{d.sh=square,d.si=0.1w}{t1}
        \fmffreeze
        \fmf{photon,left=88}{v1,v2}
        \fmfv{decoration.shape=circle,decoration.filled=empty,d.si=0.1w}{v1}
        \fmfv{decoration.shape=circle,decoration.filled=empty,d.si=0.1w}{v2}
    \end{fmfgraph*}
    }~~~,\\[24pt]
    \label{amp-d-e}
    \mathcal{A}^{(1)}_d + \mathcal{A}^{(1)}_e ~&=~~~~  
    \raisebox{-8pt}{\begin{fmfgraph*}(18,9)
        \fmfleft{i1,i2}
        \fmfright{o1,o2}
        \fmftop{t1}
        \fmf{plain}{i1,t1}
        \fmf{dbl_dashes}{t1,h2,h3,v1}
        \fmf{double}{v1,h1,v2}
        \fmf{plain}{v2,o1}
        \fmfv{label=$A$}{i1}
        \fmfv{label=$B$}{o1}
        \fmfv{d.sh=square,d.si=0.1w}{t1}
        \fmffreeze
        \fmf{photon,left=1.5}{v1,v2}
        \fmfv{decoration.shape=circle,decoration.filled=empty,d.si=0.1w}{v1}
        \fmfv{decoration.shape=circle,decoration.filled=empty,d.si=0.1w}{v2}
    \end{fmfgraph*}
    }
    ~~~~+~~~~
    \raisebox{-8pt}{\begin{fmfgraph*}(18,9)
        \fmfleft{i1,i2}
        \fmfright{o1,o2}
        \fmftop{t1}
        \fmf{plain}{i1,t1}
        \fmf{dbl_dashes}{t1,v1}
        \fmf{phantom,tension=100}{v1,h1,v2}
        \fmf{plain}{v2,o1}
        \fmfv{label=$A$}{i1}
        \fmfv{label=$B$}{o1}
        \fmfv{d.sh=square,d.si=0.1w}{t1}
        \fmffreeze
        \fmf{photon,left=88}{v1,v2}
        \fmfv{decoration.shape=circle,decoration.filled=empty,d.si=0.1w}{v1}
        \fmfv{decoration.shape=circle,decoration.filled=empty,d.si=0.1w}{v2}
    \end{fmfgraph*}
    }~~~.
\end{align}

\bigskip
\noindent Once again we encounter diagrams which are EPI but not 1PI. As discussed below, these are crucial for the cancellation of gauge dependencies. 

\vfill
\pagebreak
\section{Gauge-dependence in covariant gauge}
This section focuses on the study of gauge dependence in covariant gauges. When $\mathcal{O}$ is a neutral-current operator, the sum of the graphs discussed above are gauge invariant, whereas if $\mathcal{O}$ is a charged operator, residual gauge dependence appears in a simple factorized form that guarantees cancellation with other gauge dependent parts of the amplitude. Any ultraviolet or infrared divergences that appear below are implicitly handled with standard field theory techniques. 

In covariant (sometimes called $R_\xi$) gauges the photon propagator is given by, 
\begin{equation}
    D_{\mu\nu}(q) = \frac{-\iu}{q^2+\iu 0}\qty( g_{\mu\nu}- (1-\xi) \frac{q_\mu q_\nu}{q^2+\iu 0}) ~. 
\end{equation}
Demonstrating that amplitudes are $\xi$-independent is then equivalent to studying whether expressions with $D_{\mu\nu} \rightarrow \iu (1-\xi) q_\mu q_\nu/q^4$ vanish after summing over all graphs; this is a necessary but not sufficient condition for gauge invariance. An explicit connection to Coulomb gauge (a non-covariant gauge) is deferred to \cref{app:non-cov}.

The $\xi$-dependent part of the one-loop wavefunction renormalization is given by, 
\begin{equation}
    \frac12 Z_{A,\xi}^{(1)} = \frac{-\iu e^2}{2}(1-\xi) \int \frac{\dd \omega}{(2\pi)} \frac{\dd^3q}{(2\pi)^3} \frac{1}{q^4}\mel{A}{\rho(-\vb{q}) \rho(\vb{q})}{A}~, 
\end{equation}
where \cref{trick-1,trick-2} and the property that $q^\mu \mathscr{J}^{(\pm)}_\mu(q)=0$ have been used. 

Next, consider the vertex correction. First take diagram-($a$). Using the above identities we have, 
\begin{equation}
    \mathcal{A}_{a,\xi}^{(1)} = \iu e^2(1-\xi) \int \frac{\dd \omega}{(2\pi)} \frac{\dd^3q}{(2\pi)^3} \frac{1}{q^4}\mel{B}{\rho(-\vb{q}) \mathcal{O}(\vb{k})\rho(\vb{q})}{A}~,
\end{equation}
It is convenient to re-write this as 
\begin{equation}    
    \label{Aa-rewrite}
    \begin{split}
    \mathcal{A}_{a,\xi}^{(1)} = &~\phantom{+}\frac{\iu e^2}2(1-\xi) \int \frac{\dd \omega}{(2\pi)} \frac{\dd^3q}{(2\pi)^3} \frac{1}{q^4}\big[\mel{B}{ \mathcal{O}(\vb{k}) \rho(-\vb{q})\rho(\vb{q})}{A} +\mel{B}{\rho(-\vb{q})\rho(\vb{q}) \mathcal{O}(\vb{k}) }{A}\big] \\
    &-\frac{\iu e^2}{2}(1-\xi) \int \frac{\dd \omega}{(2\pi)} \frac{\dd^3q}{(2\pi)^3} \frac{1}{q^4}\mel{B}{[\rho(-\vb{q}),[\rho(\vb{q}), \mathcal{O}(\vb{k})] }{A} ~.
    \end{split}
\end{equation}
The commutator on the second line may be evaluated using properties of $\rho$. For operators without derivative interactions\footnote{When derivative interactions are present in $\mathcal{O}(\vb{k})$ additional seagull-like terms appear in both the commutator and the primary vertex itself (i.e., involving a photon ``coming out of the current'').} this gives \cite{Sirlin:1977sv,Weinberg:1995mt},
\begin{equation}
    \label{double-commutator}
    [\rho(-\vb{q}),[\rho(\vb{q}), \mathcal{O}(\vb{k})] = \mathcal{Q}_\mathcal{O} [\rho(-\vb{q}),\mathcal{O}(\vb{k}+\vb{q})] = Q_\mathcal{O}^2 ~\mathcal{O}(\vb{k})~,
\end{equation}
where $Q_\mathcal{O}$ is the electric charge of the operator $\mathcal{O}(\vb{k})$. Another interesting example is a the non-relativistic electromagnetic current density $\vb{J}(\vb{k})$ whose commutator gives a Schwinger term \cite{Schwinger:1959xd,Brown:1966zza,Plestid:2025ojt}, but whose double commutator vanishes. Notice that for charged current operators, this is the same gauge dependence as would be obtained from the self energy of charged point-like spectator particle of charge $Q_{\mathcal{O}}$ that is absorbed by the charged current. 

Next let us consider diagram-$(b)$. Using \cref{trick-1,trick-2} yields, 
\begin{equation}
    \label{Ab-xi-1}
    A_{b,\xi}^{(1)} = (+\iu e^2) (1-\xi) \int \frac{\dd \omega}{(2\pi)} \frac{\dd^3q}{(2\pi)^3} \frac{1}{q^4}\mel{B}{ \mathcal{O}(\vb{k}) \mathbb{P}_{\slashed{A}}  \frac{1}{-H_A+\iu 0} q^\mu J_\mu(-\vb{q})\rho(\vb{q})}{A} ~.
\end{equation}
Let us first note that $q^\mu J_\mu(\vb{q}) = \omega \rho - \vb{q}\cdot \vb{J}$, and the term proportional to $\omega$ integrates to zero. Next, it is helpful to use the identity, 
\begin{equation}
    \begin{split}
    J_i(-\vb{q})\rho(\vb{q}) &=  \frac12 [J_i(-\vb{q}) ,\rho(\vb{q}) ]+ \frac12 \{ J_i(-\vb{q}) ,\rho(\vb{q})  \} \\
    & = \frac12 q^\mu S_{\mu i}(-\vb{q},\vb{q}) + \frac12 \{ J_i(-\vb{q}) ,\rho(\vb{q})  \} ~,
    \end{split}
\end{equation}
where the second line follows after evaluating the commutator. One now sees that the gauge dependent piece related to the seagull operator cancels between $A_{b,\xi}$ and $A_{c,\xi}$ (the same happens with $A_{d,\xi}$ and $A_{e,\xi}$). This then leaves the anti-commutator contribution from $A_{b,\xi}$, 
\begin{equation}
    \label{Ab-xi-2}
    A_{\{b\},\xi }^{(1)} = \frac{\iu e^2}{2}(1-\xi) \int \frac{\dd \omega}{(2\pi)} \frac{\dd^3q}{(2\pi)^3} \frac{1}{q^4}\mel{B}{ \mathcal{O}(\vb{k}) \mathbb{P}_{\slashed{A}}  \frac{1}{-H_A+\iu 0} 
    \{\vb{q} \cdot \vb{J}(-\vb{q}),\rho(\vb{q})\}}{A} ~.
\end{equation}
Using the continuity equation $(-\vb{q}) \cdot \vb{J}(-\vb{q})= [H_A,\rho(-\vb{q})]$ and $H_A\ket{A}=0$, then gives 
\begin{equation}
    \label{Ab-xi-3}
    A_{\{b\},\xi }^{(1)} = -\frac{\iu e^2}{2}(1-\xi) \int \frac{\dd \omega}{(2\pi)} \frac{\dd^3q}{(2\pi)^3} \frac{1}{q^4}\mel{B}{ \mathcal{O}(\vb{k}) \mathbb{P}_{\slashed{A}} \rho(\vb{q})\rho(-\vb{q})}{A} ~.
\end{equation}
The wavefunction renormalization for $A$ is
\begin{equation}
    \frac12 Z_{A,\xi}^{(1)} \mathcal{A}^{(0)}= -\frac{\iu e^2}{2}(1-\xi) \int \frac{\dd \omega}{(2\pi)} \frac{\dd^3q}{(2\pi)^3} \frac{1}{q^4}\mel{B}{ \mathcal{O}(\vb{k}) \mathbb{P}_{A} \rho(\vb{q})\rho(-\vb{q})}{A}~.
\end{equation}
Adding these together then cancels the $\mathcal{O}\rho\rho$ terms on the first line of \cref{Aa-rewrite}. The same exercise can be done in the final state with the $\rho \rho \mathcal{O}$ terms, $\mathcal{A}_{\{d\},\xi}^{(1)}$ and the wavefunction renormalization for $B$. Notice that these cancellations occur for $A$ and $B$ independently. 

In summary, after adding all diagrams together the only residual gauge dependence is proportional to $\tfrac12 Q_{\mathcal{O}}^2 \mel{B}{\mathcal{O}(\vb{k})}{A}$. This residual gauge dependence for charged operators then cancels when combined with graphs involving the other charged states (e.g., an electron) produced in the reaction associated with $\mathcal{O}(\vb{k})$. The cancellations take place independently for states before and after the operator insertion $\mathcal{O}(\vb{k})$. The necessary ingredients are the cancellation of Schwinger terms (the commutator of $\vb{J}$ and $\rho$) against terms from the seagull vertex, and current conservation such that $\vb{q}\cdot \vb{J}(\vb{q})= [H,\rho(\vb{q})]$. 

\section{Isospin breaking and superallowed beta decays}
As an application of the above Feynman rules, let us consider a superallowed beta decay ($0^+\rightarrow 0^+$) mediated by the weak interaction Hamiltonian $\mathcal{H}_W$. For example, consider 
\begin{equation}
    ^{10}{\rm C}(\vb{0}) \rightarrow\! ~^{10}{\rm B}(\vb{k}) e^+(\vb{p}_e) \nu_e(\vb{p}_\nu)~,
\end{equation}
where $\vb{k}=-\vb{p}_e -\vb{p}_\nu$. At low momentum transfers, beta decays are dominantly mediated by the zeroth component of the weak charged current (the spatial component is suppressed by the mass of carbon/boron $\sim 10~{\rm GeV}$).  We therefore focus on the operator  (keeping $\vb{Q}$ as a free variable since we evaluate at $\vb{k}$ and $\vb{k}-\vb{q}$ below)
\begin{equation}
    \mathcal{O}(\vb{Q}) = \int \frac{\dd^3 p}{(2\pi)^3} b^\dagger_{\vb{p}+\vb{Q}} a_{\vb{p}}~,
\end{equation}
where $a_{\vb{p}}$ and $b^\dagger_{\vb{p}}$ annihilate protons and create neutrons respectively. When $\vb{Q}\rightarrow 0$ the operator reduces to the isospin lowering operator $\mathcal{O}(\vb{0})=\tau^-$.

Let us approximate $\mathcal{O}(\vb{k})$ by $\mathcal{O}(\vb{0})= \tau^-$.  For isovector ($T=1$) states, isospin then dictates that as $\vb{Q}\rightarrow 0$, 
\begin{align}
    \raisebox{-4pt}{\begin{fmfgraph*}(9,3)
        \fmfleft{i1,i2}
        \fmfright{o1,o2}
        \fmftop{t1}
        \fmf{plain}{i1,t1,o1}
        \fmfv{label=$A$}{i1}
        \fmfv{label=$B$}{o1}
        \fmfv{d.sh=square,d.si=0.2w}{t1}
    \end{fmfgraph*}
    }&= \sqrt{2}
\end{align}
and 
\begin{align}
        \raisebox{-4pt}{\begin{fmfgraph*}(9,3)
        \fmfleft{i1,i2}
        \fmfright{o1,o2}
        \fmftop{t1}
        \fmf{plain}{i1,t1}
        \fmf{dbl_dashes}{t1,o1}
        \fmfv{d.sh=square,d.si=0.2w}{t1}
    \end{fmfgraph*}
    }
    &= 
    \raisebox{-4pt}{\begin{fmfgraph*}(9,3)
        \fmfleft{i1,i2}
        \fmfright{o1,o2}
        \fmftop{t1}
        \fmf{plain}{t1,o1}
        \fmf{dbl_dashes}{i1,t1}
        \fmfv{d.sh=square,d.si=0.2w}{t1}
    \end{fmfgraph*}
    }
    =0
\end{align}
These relations simplify some of the following analysis. For example the diagrams $(b)-(e)$ from \cref{amp-b-c,amp-d-e} do not contribute to the amplitude because of the vanishing Feynman rule given above. Although we have $\vb{k}=0$ in mind, we retain $\mathcal{O}(\vb{k})$ in much of the discussion below, which makes generalization to larger momentum transfers straightforward. 

Radiative corrections to beta decay can be broken apart into: the wavefunction renormalization of the electron, $Z_e$,  the nuclear vertex correction, $V$, and everything else which we will refer to as ``box graphs'' $B$. Next, define the radiative corrections multiplicatively, 
\begin{equation}
    (B e^+ \nu_e|\mathcal{H}_W|A) = \qty[1 + V + B + (\sqrt{Z_e} -1) ]\times \qty(~~~\raisebox{-15pt}{\begin{fmfgraph*}(9,9)
        \fmfleft{i0,i1,l1,l2,l3,l4,l5,l6,i2,i3}
        \fmfright{o0,o1,k1,k2,k3,k4,k5,k6,o2,o3}
        \fmf{plain}{i2,p1,v1,v2,o2}
        \fmf{plain}{i1,v3}
        \fmf{plain}{v1,v3}
        \fmf{plain}{v1,p2,o1}
        \fmfv{label=$A$}{l1}
        \fmfv{label=$B$}{k1}
        \fmfv{label=$\nu$}{l6}
         \fmfv{label=$e$}{k6}
        \fmfv{d.sh=square,d.si=0.15w}{v1}
        \fmffreeze
    \end{fmfgraph*}
    }~~~)~.
\end{equation}
The vertex correction may be defined as $(B|\mathcal{O}(\vb{k})|A) = (1+V) \times \mel{B}{\mathcal{O}(\vb{k})}{A}$ and is the object discussed extensively in the preceding sections (see also Refs.~\cite{Preparata:1968ioj,Sirlin:1977sv,Seng:2021syx}). 
With superscripts denoting orders in $\alpha$, through one-loop order this formula reads, 
\begin{equation}
    \label{RC-decomp-1}
    (B e^+ \nu_e|\mathcal{H}_W|A) = \qty[1 +V^{(1)} + B^{(1)} + \frac12 Z_e^{(1)}] \times\qty(~~~\raisebox{-15pt}{\begin{fmfgraph*}(9,9)
        \fmfleft{i0,i1,l1,l2,l3,l4,l5,l6,i2,i3}
        \fmfright{o0,o1,k1,k2,k3,k4,k5,k6,o2,o3}
        \fmf{plain}{i2,p1,v1,v2,o2}
        \fmf{plain}{i1,v3}
        \fmf{plain}{v1,v3}
        \fmf{plain}{v1,p2,o1}
        \fmfv{label=$A$}{l1}
        \fmfv{label=$B$}{k1}
        \fmfv{label=$\nu$}{l6}
         \fmfv{label=$e$}{k6}
        \fmfv{d.sh=square,d.si=0.15w}{v1}
        \fmffreeze
    \end{fmfgraph*}
    }~~~)  ~. 
\end{equation}
Diagrammatically the one-loop expressions are given by (using $\bullet=(-1)$ again),
\begin{align}
    Z_e^{(1)}&=  ~~~~~\raisebox{-22pt}{\begin{fmfgraph*}(14,9)
        \fmfleft{i1}
        \fmfright{o1}
        \fmf{plain}{i1,v1,v2,v3,v4,v5,o1}
         \fmfv{label=$e$}{i1,o1}
        \fmfv{d.sh=circle,d.si=0.075w}{v3}
        \fmffreeze
        \fmf{photon,left=1}{v1,v5}
    \end{fmfgraph*}
    }~~~~~~~,\\
    V^{(1)}&=  \left(\rule{0cm}{1.25cm}\right.~~
     \raisebox{-14pt}{\begin{fmfgraph*}(12,8)
        \fmfleft{i1,i2}
        \fmfright{o1,o2}
        \fmf{plain}{i2,p1,v1,p2,o2}
        \fmf{plain}{i1,v2}
        \fmf{double}{v2,v1,v3}
        \fmf{plain}{v3,o1}
        \fmfv{label=$A$}{i1}
        \fmfv{label=$B$}{o1}
        \fmfv{label=$\nu$}{i2}
         \fmfv{label=$e$}{o2}
        \fmfv{d.sh=square,d.si=0.1w}{v1}
        \fmffreeze
        \fmf{photon,right=1}{v2,v3}
        \fmfv{decoration.shape=circle,decoration.filled=empty,d.si=0.1w}{v3}
        \fmfv{decoration.shape=circle,decoration.filled=empty,d.si=0.1w}{v2}
    \end{fmfgraph*}
    }
    ~
    \left. \rule{0cm}{1.25cm}\right)
    \times 
    \left( \rule{0cm}{1.25cm}\right.~~
    \raisebox{-14pt}{\begin{fmfgraph*}(12,8)
        \fmfleft{i1,i2}
        \fmfright{o1,o2}
        \fmf{plain}{i2,p1,v1,v2,o2}
        \fmf{plain}{i1,p2,v1}
        \fmf{plain}{v1,v3}
        \fmf{plain}{v3,o1}
        \fmfv{label=$A$}{i1}
        \fmfv{label=$B$}{o1}
        \fmfv{label=$\nu$}{i2}
         \fmfv{label=$e$}{o2}
        \fmfv{d.sh=square,d.si=0.1w}{v1}
        \fmffreeze
    \end{fmfgraph*}
    }
    ~\left. \rule{0cm}{1.25cm}\right)^{-1}
    \!\!\!+~
    \frac12~\qty(~~\raisebox{-28pt}{\begin{fmfgraph*}(12,12)
        \fmfleft{i1}
        \fmfright{o1} 
        \fmf{plain}{i1,p1,v1}
        \fmf{double}{v1,v3,v5,v4,v2}
        \fmf{plain}{v2,p2,o1}
        \fmfv{label=$A$}{p1,p2}
        \fmfdot{v5}
        \fmfv{decoration.shape=circle,decoration.size=0.1w,decoration.filled=empty}{v1,v2}
        \fmffreeze
        \fmf{photon,left=1}{v1,v2}
        %\fmf{photon,left=2}{v1,v2}
    \end{fmfgraph*}
    } 
    ~~~+~~~
    \raisebox{-28pt}{\begin{fmfgraph*}(12,12)
        \fmfleft{i1}
        \fmfright{o1} 
        \fmf{plain}{i1,p1,v1}
        \fmf{double}{v1,v3,v5,v4,v2}
        \fmf{plain}{v2,p2,o1}
        \fmfv{label=$B$}{p1,p2}
        \fmfdot{v5} 
        \fmfv{decoration.shape=circle,decoration.size=0.1w,decoration.filled=empty}{v1,v2}
        \fmffreeze
        \fmf{photon,left=1}{v1,v2}
        %\fmf{photon,left=2}{v1,v2}
    \end{fmfgraph*}
    }~ )~,
    \\
    B^{(1)}&=  \left(\rule{0cm}{1.25cm}\right.~~~
    \raisebox{-14pt}{\begin{fmfgraph*}(12,8)
        \fmfleft{i1,i2}
        \fmfright{o1,o2}
        \fmf{plain}{i2,p1,v1,v2,o2}
        \fmf{plain}{i1,v3}
        \fmf{double}{v1,v3}
        \fmf{plain}{v1,p2,o1}
        \fmfv{label=$A$}{i1}
        \fmfv{label=$B$}{o1}
        \fmfv{label=$\nu$}{i2}
         \fmfv{label=$e$}{o2}
        \fmfv{d.sh=square,d.si=0.1w}{v1}
        \fmffreeze
        \fmf{photon,right=1}{v2,v3}
        \fmfv{decoration.shape=circle,decoration.filled=empty,d.si=0.1w}{v3}
    \end{fmfgraph*}
    }
    ~~~
    +
    ~~~
    \raisebox{-14pt}{\begin{fmfgraph*}(12,8)
        \fmfleft{i1,i2}
        \fmfright{o1,o2}
        \fmf{plain}{i2,p1,v1,v2,o2}
        \fmf{plain}{i1,p2,v1}
        \fmf{double}{v1,v3}
        \fmf{plain}{v3,o1}
        \fmfv{label=$A$}{i1}
        \fmfv{label=$B$}{o1}
        \fmfv{label=$\nu$}{i2}
         \fmfv{label=$e$}{o2}
        \fmfv{d.sh=square,d.si=0.1w}{v1}
        \fmffreeze
        \fmf{photon,left=1}{v2,v3}
        \fmfv{decoration.shape=circle,decoration.filled=empty,d.si=0.1w}{v3}
    \end{fmfgraph*}
    }~~~
    \left. \rule{0cm}{1.25cm}\right)
    \times 
 \left( \rule{0cm}{1.25cm}\right.~~
 \raisebox{-14pt}{\begin{fmfgraph*}(12,8)
        \fmfleft{i1,i2}
        \fmfright{o1,o2}
        \fmf{plain}{i2,p1,v1,v2,o2}
        \fmf{plain}{i1,p2,v1}
        \fmf{plain}{v1,v3}
        \fmf{plain}{v3,o1}
        \fmfv{label=$A$}{i1}
        \fmfv{label=$B$}{o1}
        \fmfv{label=$\nu$}{i2}
         \fmfv{label=$e$}{o2}
        \fmfv{d.sh=square,d.si=0.1w}{v1}
        \fmffreeze
    \end{fmfgraph*}
    }
~\left. \rule{0cm}{1.25cm}\right)^{-1}
    ~.
\end{align}
Each three of these terms is gauge dependent, and so it proves convenient to add and subtract the following term: $\qty(\frac12 Z_p^{(1)} + B_p^{(1)}) \times \mel{Be^+\nu_e}{\mathcal{H}_W}{A}$, where $Z_p^{(1)}$ is the wavefunction renormalization of a heavy point-like particle of charge one (i.e., a  proton), 
\begin{align}
    \begin{split}
    Z_p^{(1)} &= - e^2 \int \frac{\dd^4 q}{(2\pi)^4}  D^{\mu\nu}(q) v_\mu  \qty(\frac{1}{\omega +\iu 0})^2 v_\nu ~\\[6pt]
    &= ~~~~~\raisebox{-22pt}{\begin{fmfgraph*}(14,9)
        \fmfleft{i1}
        \fmfright{o1}
        \fmf{plain}{i1,v1,v2,v3,v4,v5,o1}
         \fmfv{label=$p$}{i1,o1}
        \fmfv{d.sh=circle,d.si=0.075w}{v3}
        \fmffreeze
        \fmf{photon,left=1}{v1,v5}
    \end{fmfgraph*}
    }~~~~~~. 
    \end{split}
\end{align}
with $\bullet=(-1)$, and $B_p^{(1)}$ is related to the box diagram for a heavy point-like proton, 
\begin{align}
    B_p^{(1)}  &= \frac{1}{\bar{u}(p_\nu)\Gamma v(p_e)}\times e^2 \int \frac{\dd^4 q}{(2\pi)^4}  D^{\mu\nu}(q) v_\mu  \qty(\frac{1}{\omega +\iu 0}) \bar{u}(p_\nu) \Gamma \frac{1}{-\slashed{p}_e-\slashed{q}-m} \gamma_\nu v(p_e) ~,
\end{align}
with $\Gamma=(1-\gamma_5)\slashed{v}$ the weak vertex for the vector current. Diagrammatically this can be written as 
\begin{equation}
     B_p^{(1)} =    \left(\rule{0cm}{1.25cm}\right.~~~
    \raisebox{-14pt}{\begin{fmfgraph*}(12,8)
        \fmfleft{i1,i2}
        \fmfright{o1,o2}
        \fmf{plain}{i2,p1,v1,v2,o2}
        \fmf{plain}{i1,v3}
        \fmf{plain}{v1,v3}
        \fmf{plain}{v1,p2,o1}
        \fmfv{label=$p$}{i1}
        \fmfv{label=$n$}{o1}
        \fmfv{label=$\nu$}{i2}
         \fmfv{label=$e$}{o2}
        \fmfv{d.sh=square,d.filled=empty,d.si=0.1w}{v1}
        \fmffreeze
        \fmf{photon,right=1}{v2,v3}~
    \end{fmfgraph*}
    }
    ~~
    \left. \rule{0cm}{1.25cm}\right)
    \times 
 \left( \rule{0cm}{1.25cm}\right.~~
 \raisebox{-14pt}{\begin{fmfgraph*}(12,8)
        \fmfleft{i1,i2}
        \fmfright{o1,o2}
        \fmf{plain}{i2,p1,v1,v2,o2}
        \fmf{plain}{i1,p2,v1}
        \fmf{plain}{v1,v3}
        \fmf{plain}{v3,o1}
        \fmfv{label=$p$}{i1}
        \fmfv{label=$n$}{o1}
        \fmfv{label=$\nu$}{i2}
         \fmfv{label=$e$}{o2}
        \fmfv{d.sh=square,d.filled=empty,d.si=0.1w}{v1}
        \fmffreeze
    \end{fmfgraph*}
    }
~\left. \rule{0cm}{1.25cm}\right)^{-1}~,
\end{equation}
where the empty square denotes the point-like weak interaction vertex.

Defining the subtracted vertex correction and box diagrams, 
\begin{align}
   \tilde{V}^{(1)} &= V^{(1)} - \frac12 Z_p^{(1)}~, \\
   \tilde{B}^{(1)} &= B^{(1)} - B_p^{(1)}~,  
\end{align}
and the virtual part of the ``outer terms'' (to match onto historical nomenclature \cite{Sirlin:1967zza,Hardy:2020qwl})
\begin{equation}
    \tilde{O}_{\rm virt.}^{(1)} =  \left(\rule{0cm}{1.25cm}\right.~~
    \raisebox{-14pt}{\begin{fmfgraph*}(12,8)
        \fmfleft{i1,i2}
        \fmfright{o1,o2}
        \fmf{plain}{i2,p1,v1,v2,o2}
        \fmf{plain}{i1,v3}
        \fmf{plain}{v1,v3}
        \fmf{plain}{v1,p2,o1}
        \fmfv{label=$p$}{i1}
        \fmfv{label=$n$}{o1}
        \fmfv{label=$\nu$}{i2}
         \fmfv{label=$e$}{o2}
        \fmfv{d.sh=square,d.filled=empty,d.si=0.1w}{v1}
        \fmffreeze
        \fmf{photon,right=1}{v2,v3}~
    \end{fmfgraph*}
    }
    ~
    \left. \rule{0cm}{1.25cm}\right)
    \times 
 \left( \rule{0cm}{1.25cm}\right.~~
 \raisebox{-14pt}{\begin{fmfgraph*}(12,8)
        \fmfleft{i1,i2}
        \fmfright{o1,o2}
        \fmf{plain}{i2,p1,v1,v2,o2}
        \fmf{plain}{i1,p2,v1}
        \fmf{plain}{v1,v3}
        \fmf{plain}{v3,o1}
        \fmfv{label=$p$}{i1}
        \fmfv{label=$n$}{o1}
        \fmfv{label=$\nu$}{i2}
         \fmfv{label=$e$}{o2}
        \fmfv{d.sh=square,d.filled=empty,d.si=0.1w}{v1}
        \fmffreeze
    \end{fmfgraph*}
    }
    ~\left. \rule{0cm}{1.25cm}\right)^{-1}\!
    + 
    ~\frac12~\qty(~~\raisebox{-22pt}{\begin{fmfgraph*}(14,9)
        \fmfleft{i1}
        \fmfright{o1}
        \fmf{plain}{i1,v1,v2,v3,v4,v5,o1}
         \fmfv{label=$\!\!p$}{o1}
         \fmfv{label=$p\!\!$}{i1}
        \fmfv{d.sh=circle,d.si=0.1w}{v3}
        \fmffreeze
        \fmf{photon,left=1}{v1,v5}
    \end{fmfgraph*}
    }
    ~~~+ ~~~
    \raisebox{-22pt}{\begin{fmfgraph*}(14,9)
        \fmfleft{i1}
        \fmfright{o1}
        \fmf{plain}{i1,v1,v2,v3,v4,v5,o1}
         \fmfv{label=$\!\!e$}{o1}
         \fmfv{label=$e\!\!$}{i1}
        \fmfv{d.sh=circle,d.si=0.1w}{v3}
        \fmffreeze
        \fmf{photon,left=1}{v1,v5}
    \end{fmfgraph*}
    }
    ~)~,
\end{equation}
we then re-write \cref{RC-decomp-1} as 
\begin{equation}
    \label{RC-decomp-2}
    (B e^+ \nu_e|\mathcal{H}_W|A) = \qty[1 +\tilde{V}^{(1)} + \tilde{B}^{(1)} + \tilde{O}_{\rm virt.}^{(1)}] \times\qty(~~~\raisebox{-15pt}{\begin{fmfgraph*}(9,9)
        \fmfleft{i0,i1,l1,l2,l3,l4,l5,l6,i2,i3}
        \fmfright{o0,o1,k1,k2,k3,k4,k5,k6,o2,o3}
        \fmf{plain}{i2,p1,v1,v2,o2}
        \fmf{plain}{i1,v3}
        \fmf{plain}{v1,v3}
        \fmf{plain}{v1,p2,o1}
        \fmfv{label=$A$}{l1}
        \fmfv{label=$B$}{k1}
        \fmfv{label=$\nu$}{l6}
         \fmfv{label=$e$}{k6}
        \fmfv{d.sh=square,d.si=0.15w}{v1}
        \fmffreeze
    \end{fmfgraph*}
    }~~~)  ~. 
\end{equation}
The quantities in the above expression are closely related to existing conventions in the literature for the nuclear structure dependent correction\footnote{The subtracted box $\tilde{B}^{(1)}$ contains the Fermi function, shape factor, and $\delta_{\rm NS}$.} $\delta_{\rm NS}$ or ``the $\gamma W$ box'', and the virtual part of the outer radiative corrections $\delta_{\rm R}'$ (i.e., without real photon emission).  Our focus is on the vertex correction $\tilde{V}^{(1)}$.

The decomposition in \cref{RC-decomp-2} has a  number of advantages as compared to \cref{RC-decomp-1}. First, notice that $O^{(1)}_{\rm virt.}$ is gauge invariant, being the complete set of diagrams for the crossed reaction of $\bar{\nu}_e p^+ \rightarrow e^+ n$ (inverse beta decay). Therefore the sum $\tilde{B}^{(1)} + \tilde{V}^{(1)}$ is also gauge invariant. Second, using the results presented above, it is easy to show that $\tilde{V}^{(1)}$ is independent of the gauge parameter $\xi$ in covariant gauges. Similarly, using the results from Ref.~\cite{Plestid:2025ojt} one can show that 
\begin{equation}
    \begin{split}
    \tilde{B}^{(1)} = &\frac{1}{\bar{u}(p_\nu)\Gamma v(p_e)\mel{B}{\mathcal{O}(\vb{k})}{A}} \times e^2 \int \frac{\dd^4q}{(2\pi)^4 } D^{\mu\nu}(q) \bar{u}(p_\nu) \Gamma\frac{1}{-\slashed{p}_e-\slashed{q}-m} \gamma_\nu v(p_e)  \\
    & \hspace{0.0\linewidth}\times\bigg[ -(2\pi \iu) v_\mu \delta(\omega) \mel{B}{\rho(\vb{q})\mathcal{O}(\vb{k}-\vb{q})}{A} + \\
     & \hspace{0.05\linewidth} \mel{B}{\mathscr{J}_\mu^{(-)}(\vb{q}) \frac{1}{-\omega-H_B}\mathcal{O}(\vb{k}-\vb{q})}{A}  + \mel{B}{\mathcal{O}(\vb{k}-\vb{q}) \frac{1}{\omega-H_A}\mathscr{J}_\mu^{(+)}(\vb{q})}{A}\bigg ]~,
     \end{split}
\end{equation}
and since the term in square brackets, $H_\mu$, satisfies $q^\mu H_\mu=0$ one can again explicitly show that $\tilde{B}^{(1)}$ is $\xi$-independent in covariant gauge. Nevertheless, as is discussed in detail in \cref{app:non-cov}, $\tilde{B}^{(1)}$ and $\tilde{V}^{(1)}$ are both gauge dependent (as can be seen explicitly by comparing non-covariant gauges). For example the difference between $\tilde{V}^{(1)}_F$ computed in Feynman gauge, and $\tilde{V}^{(1)}_C$ computed in Coulomb gauge is 
\begin{equation}
    \label{V1-gauge-diff}
    \begin{split}
      &\qty[\tilde{V}^{(1)}_F-\tilde{V}^{(1)}_C] = -\iu e^2 \int \frac{\dd^4 q}{(2\pi)^4}\frac{1}{q^2}\frac{1}{\omega+\iu 0} \frac{\mel{B}{O(\vb{k},\vb{q},\omega)}{A}}{\mel{B}{\mathcal{O}(\vb{k})}{A}} ~,
    \end{split}
\end{equation}
where (note that $\omega\rightarrow -|\vb{q}|$ after closing the energy contour in the upper half plane), 
\begin{equation}
    O(\vb{k},\vb{q},\omega)=\mathcal{O}(\vb{k}-\vb{q})\frac{1}{\omega-H_A+\iu 0} \rho(\vb{q})-\rho(\vb{q})\frac{1}{\omega-H_B+\iu 0}  \mathcal{O}(\vb{k}-\vb{q})- \frac{1}{\omega+\iu 0}\mathcal{O}(\vb{k})~.
\end{equation}
We estimate that $\qty[\tilde{V}^{(1)}_F-\tilde{V}^{(1)}_C] <(\alpha/\pi)$. This difference may be added to $\tilde{V}^{(1)}_C$ (computed in Coulomb gauge) to obtain the answer in Feynman gauge (and thereby the whole class of covariant gauges), 
\begin{align}
    \label{V1-C}
        \tilde{V}^{(1)}_C =&\frac{e^2}{\mel{B}{\mathcal{O}(\vb{k})}{A}}\int  
        \frac{\dd^3 q}{(2\pi)^3}\frac{1}{2|\vb{q}|} \mel{B}{J_i(-\vb{q})\frac{1}{|\vb{q}|+H_B}\mathcal{O}(\vb{k}) \frac{1}{|\vb{q}|+H_A} J_j(\vb{q})}{A} \mathbb{P}_{ij}(\vb{q}) \nonumber \\ 
        &\hspace{0.1\linewidth}-\frac12e^2 \int \frac{\dd^3 q}{(2\pi)^3}\frac{1}{2|\vb{q}|} \mel{A}{J_i(-\vb{q})\qty(\frac{1}{|\vb{q}|+H_A})^2 J_j(\vb{q})}{A} \mathbb{P}_{ij}(\vb{q})\\
         &\hspace{0.1\linewidth}-\frac12e^2 \int \frac{\dd^3 q}{(2\pi)^3}\frac{1}{2|\vb{q}|} \mel{B}{J_i(-\vb{q})\qty(\frac{1}{|\vb{q}|+H_B})^2 J_j(\vb{q})}{B} \mathbb{P}_{ij}(\vb{q})~, \nonumber
\end{align}
where the infinitesimal in $1/(|\vb{q}|+H_{A,B}-\iu0)$ has been left implicit and $\mathbb{P}_{ij}(\vb{q})= \delta_{ij} - q_i q_j/\vb{q}^2$ is a transverse projector. When $\vb{k}=0$ (and when isospin is a good symmetry of the strong interaction) this can be further simplified to a two-point function between $J_i$ and the charged weak-current $\mathcal{J}^-_i(\vb{q})=[\tau^-,J_i(\vb{q})]$ \cite{Sirlin:1977sv} which is analogous to \cref{V1-gauge-diff} (see also discussion in Refs.~\cite{Preparata:1968ioj,Sirlin:1977sv,Seng:2021syx}). This is most easily seen by noting that $\tau^-$ commutes with $H_{A,B}$ and using $\tau^-\ket{A} = \sqrt{2}\ket{B}$
\begin{align}
    \label{V1a-C}
        \tilde{V}^{(1)}_C =&-\frac{e^2}{2}\int  
        \frac{\dd^3 q}{(2\pi)^3}\frac{1}{2|\vb{q}|} \mel{B}{\mathcal{J}^-_i(-\vb{q})\qty(\frac{1}{|\vb{q}|+H_B})^2 J_j(\vb{q})}{A} \mathbb{P}_{ij}(\vb{q}) \nonumber \\
         &+\frac{e^2}{2}\int  
        \frac{\dd^3 q}{(2\pi)^3}\frac{1}{2|\vb{q}|} \mel{B}{J_i(-\vb{q})\qty(\frac{1}{|\vb{q}|+H_B})^2\mathcal{J}_j^-(\vb{q})}{A} \mathbb{P}_{ij}(\vb{q})~.
\end{align}
We estimate that $\tilde{V}^{(1)}_C\sim (\alpha/\pi) v_p^2 \sim 1\times 10^{-4} $ with $v_p$ a typical proton velocity, (there are $Z$-enhancements term-by-term in \cref{V1-C}, but they cancel as can be seen in \cref{V1a-C}). 

Let us now discuss the mapping of our $\tilde{V}^{(1)}$ and $\tilde{B}^{(1)}$ onto conventions in the literature (namely for $\delta_C$, $\delta_{\rm NS}$, and $\delta_R'$). These quantities are typically defined via \cite{Hardy:2020qwl,Seng:2022cnq}
\begin{equation}
    \Gamma = \qty[\Gamma]_{\substack{\text{~}\\\text{leading}\\\text{in $Z\alpha$}}} ~\times ~(1-\delta_{\rm C}+\delta_{\rm NS})(1+\delta_{\rm R}')~,
\end{equation}
where we omit the short-distance ``inner correction'' $\Delta_{\rm R}$ assuming it has already been subsumed into the tree-level matrix element. The leading-in-$Z\alpha$ piece includes all terms which can be obtained by solving for the Dirac wavefunction of the positron. 

The virtual part of the outer correction, $\delta_{\rm R}'$, is entirely contained within $\mathcal{O}_{\rm virt.}^{(1)}$ and we do not discuss it further. Historically the isospin correction $(1-\delta_C/2)$ is defined as the wavefunction overlap between the initial and final state with the Coulomb force included $(1-\delta_C/2)\equiv (B|A)$. We are able to identify this effect as at two-loops in our formalism, and comment on this contribution below. The sum $\tilde{V}^{(1)}+\tilde{B}^{(1)}$ includes $\delta_{\rm NS}$ as well as some of the leading-in-$Z\alpha$ effects. For example, the product of the  Fermi function, $F(Z,E)$, and shape factor $C(Z,E)$,  are also contained within this quantity (see \cref{app:delta-NS} for more details). 

At higher loop order we can make contact with existing studies of the isospin breaking \cite{Miller:2008my,Miller:2009cg} correction $\delta_C$, even in the absence of a detailed understanding of gauge dependencies at two-loop order. The largest source of isospin breaking stems from the intranuclear Coulomb interaction, which be identified by keeping all terms at any order in perturbation theory and replacing $J_\mu \rightarrow v_\mu \rho$. These contributions must be gauge invariant, since they are maximally $Z$-enhanced and unsuppressed by proton-velocities in the non-relativistic limit. 

Keeping only these ``Coulomb terms'' we find that the maximally $Z$-enhanced parts from $V$ match onto the existing definition of $\delta_{\rm C}$ in the literature. For example at $O(Z^2 \alpha^2)$, taking the following diagrams (vertex correction and wavefunction renormalization),
\begin{equation}
    \raisebox{-8pt}{\begin{fmfgraph*}(18,9)
        \fmfleft{i1,i2}
        \fmfright{o1,o2}
        \fmftop{t1}
        \fmf{plain}{i1,v1}
        \fmf{double}{v1,h1,v2}
        \fmf{dbl_dashes}{v2,h2,h3,t1}
        \fmf{dbl_dashes}{t1,h4,h5,v3}
        \fmf{double}{v3,h6,v4}
        \fmf{plain}{v4,o1}
        \fmfv{label=$A$}{i1}
        \fmfv{label=$B$}{o1}
        \fmfv{d.sh=square,d.si=0.1w}{t1}
        \fmffreeze
        \fmf{photon,left=1.5}{v1,v2}
        \fmf{photon,left=1.5}{v3,v4} 
        \fmfv{decoration.shape=circle,decoration.filled=empty,d.si=0.1w}{v1}
        \fmfv{decoration.shape=circle,decoration.filled=empty,d.si=0.1w}{v2}
        \fmfv{decoration.shape=circle,decoration.filled=empty,d.si=0.1w}{v3}
        \fmfv{decoration.shape=circle,decoration.filled=empty,d.si=0.1w}{v4}
    \end{fmfgraph*}
    }
    \quad
    \quad , 
    \quad
    \quad
    \raisebox{8pt}{\begin{fmfgraph*}(20,10)
        \fmfleft{i1,i2}
        \fmfright{o1,o2} 
        \fmf{phantom}{i1,o2}
        \fmf{plain}{i1,v1}
        \fmf{double}{v1,h1,v2}
        \fmf{dbl_dashes}{v2,h7,h8,v3}
        \fmf{double}{v3,h2,v4}
        \fmf{plain}{v4,o1} 
        \fmfv{label=$A$}{i1}
        \fmfv{label=$A$}{o1}
        \fmffreeze
        \fmf{photon,left=1.5}{v1,v2}
        \fmf{photon,left=1.5}{v3,v4} 
        \fmfv{decoration.shape=circle,decoration.filled=empty,d.si=0.1w}{v1}
        \fmfv{decoration.shape=circle,decoration.filled=empty,d.si=0.1w}{v2}
        \fmfv{decoration.shape=circle,decoration.filled=empty,d.si=0.1w}{v3}
        \fmfv{decoration.shape=circle,decoration.filled=empty,d.si=0.1w}{v4}
    \end{fmfgraph*}
    }
    \quad
    \quad , 
    \quad
    \quad
    \raisebox{8pt}{\begin{fmfgraph*}(20,10)
        \fmfleft{i1,i2}
        \fmfright{o1,o2} 
        \fmf{phantom}{i1,o2}
        \fmf{plain}{i1,v1}
        \fmf{double}{v1,h1,v2}
        \fmf{dbl_dashes}{v2,h7,h8,v3}
        \fmf{double}{v3,h2,v4}
        \fmf{plain}{v4,o1} 
        \fmfv{label=$B$}{i1}
        \fmfv{label=$B$}{o1}
        \fmffreeze
        \fmf{photon,left=1.5}{v1,v2}
        \fmf{photon,left=1.5}{v3,v4} 
        \fmfv{decoration.shape=circle,decoration.filled=empty,d.si=0.1w}{v1}
        \fmfv{decoration.shape=circle,decoration.filled=empty,d.si=0.1w}{v2}
        \fmfv{decoration.shape=circle,decoration.filled=empty,d.si=0.1w}{v3}
        \fmfv{decoration.shape=circle,decoration.filled=empty,d.si=0.1w}{v4}
    \end{fmfgraph*}
    } \quad , 
\end{equation}
\vspace{6pt}

\noindent and replacing $J_\mu \rightarrow v_\mu \rho$ then reproduces the $O(Z^2\alpha^2)$ expressions for isospin breaking derived in Refs.~\cite{Miller:2008my,Miller:2009cg} using time-independent perturbation theory. More explicitly, the vertex diagram corresponds to the second term in Eq.~(19) of Ref.~\cite{Miller:2008my}, and the two wavefunction renormalization diagrams map onto the factor of $\sqrt{Z_i}\sqrt{Z_f}$ in the same equation of that reference.

The results presented here for the nuclear vertex correction $V^{(1)}$, and its subtracted counterpart $\tilde{V}^{(1)}$, generalize the analysis of isospin breaking to include transverse photons and to account for a subtle gauge dependence. This gauge dependence is properly handled in existing dispersive treatments. Our discussion of $\mathcal{O}_{\rm virt.}^{(1)}$ provides a novel derivation of the (well known) factorization of outer radiative corrections $\delta_{R}'$ from the tree level matrix element $\mel{B}{\mathcal{O}(\vb{k})}{A}$ at $O(\alpha)$ \cite{Sirlin:1967zza}.

\pagebreak

We have therefore reproduced  well known results at one-loop order derived using Brown's on-shell perturbation formula (as opposed to the LSZ construction described herein). Moreover we have also identified $\delta_C$ (as well as the outer corrections, Fermi function, and shape factor) in the same formalism. We therefore believe that the LSZ construction presented here contains {\it all } electromagnetically induced corrections and can be systematically extended to higher orders in perturbation theory. Such higher order corrections will bear on the ongoing efforts to extract  $|V_{ud}|$ and to search for new physics.

\section{Conclusions}
In this paper we have studied QED corrections to the matrix elements of external operators between composite states. The formalism can be applied at arbitrary loop order using a standard diagrammatic expansion. These corrections are important for any processes involving atoms, nuclei, or heavy hadrons in which percent-level control of the theory is required. Relevant examples include neutrino nucleus scattering, radiative and non-radiative $B$-decays, and superallowed beta decay. 

 Our LSZ construction is somewhat unconventional since we have allowed ourselves the freedom to separate the Hilbert space of states into two arbitrary disjoint sectors rather than 1PI and 1PR graphs. The simplest choice is to omit only the external particle in the irreducible graphs (EPI). We have also provided an explicit discussion of how the LSZ reduction relates to time-independent perturbation theory. 

The perturbative expansion can be summarized with operator valued Feynman rules including projectors that are used to define EPI graphs. The graphical structure of the vertex correction, and wavefunction renormalization are similar, but not identical to scalar-QED. The differences stem from the fact that the strong-interaction propagator includes both single particle states that are not EPI, and also continuum contributions that would correspond to loop diagrams (e.g., from an emitted and reabsorbed neutron) if the strong interaction were treated perturbatively. Using the Feynman rules defined above, it is a simple matter of drawing graphs, and transcribing the result in terms of on-shell matrix elements $\mel{B}{\ldots}{A}$. 

\pagebreak

Motivated by our recent work~\cite{Plestid:2025ojt}, one of our goals has been to understand how gauge dependence appears and ultimately cancels in physical observables. For neutral operators we have shown\footnote{We have not explicitly analyzed derivative-operators, but an extension to this case is straightforward. } that the wavefunction renormalization and vertex corrections combined together are gauge invariant. For charged current operators, the commutator $[\rho,\mathcal{O}]$ does not vanish, and one picks up additional gauge dependence. In covariant, or $R_\xi$, gauges the $\xi$-dependence appears from a double-commutator. In non-covariant gauges, such as Coulomb gauge, the result is more subtle. 

We have applied our analysis to nuclear beta decays, with a specific eye towards the high precision goals necessary to test first-row CKM-unitarity. We have the following results: 
\begin{itemize}
    \item All QED effects (the Fermi function, isospin breaking, nuclear structure corrections, and the factorization of outer corrections) are described in a single formalism. 
    \item The LSZ formalism reproduces all known one-loop effects, and the isospin breaking correction $\delta_C$ at two-loop order. Furthermore it can be readily extended to higher orders in perturbation theory e.g., two-loop $Z\alpha^2$ corrections.
    \item A detailed understanding of gauge dependencies in formalism.  These details are useful when constructing approximation schemes \cite{Plestid:2025ojt}.
\end{itemize}

There are a variety of phenomenological applications that should be pursued. Radiative corrections to neutrino nucleus scattering in the $0.5-2.5~{\rm GeV}$ regime are important for the long-baseline neutrino oscillation program. Since we have retained momentum dependence in $\mathcal{O}(\vb{k})$ above, many of our results can be applied to this problem.   Another interesting application is neutral current scattering (such as electron scattering, coherent neutrino nucleus scattering, and parity violating electron scattering), where the hadronic vertex corrections derived in this paper become gauge invariant. Finally, it would be interesting to study structure dependent radiative corrections that appear in the theory of $B$ decays. 

In summary, we have derived operator-valued Feynman rules for computing perturbative QED corrections to processes involving composite bound states.\!\footnotemark[1] Notable developments include the notion of EPI vs 1PI, and connections to Brillouin-Wigner perturbation theory. We have developed a set of simple Feynman rules that allow for systematically improvable QED calculations that are relevant for extractions of the CKM  matrix element $|V_{ud}|$.  There are many opportunities for the phenomenological application of these results that bear on important questions in high energy and nuclear physics.

\section*{Acknowledgments} 
\vspace{-12pt}
  We thank Chien-Yeah Seng for a detailed discussion on the relation between our results and the existing one-loop literature, and Vincenzo Cirigliano for reading, and giving helpful feedback on, this manuscript.  RP thanks FRIB and the theory groups at INFN Frascati and Fermilab for their hospitality. RP and MBW are supported by the U.S. Department of Energy under Award Number DE-SC0011632, and by the Walter Burke Institute for Theoretical Physics. RP is supported by the Neutrino Theory Network under Award Number DEAC02-07CH11359.

\appendix

\section{\label{app:non-cov} Relation between Feynman and Coulomb gauge} 
In the main text we have demonstrated the equivalence of all $R_\xi$ gauges, however we find that results are simplest in Coulomb gauge (which is not an $R_\xi$ or covariant gauge). In this appendix we begin with the result in Feynman gauge and demonstrate equivalence with Coulomb gauge explicitly. A more general discussion of non-covariant gauges should be pursued in future work. 

Consider the propagator in Coulomb gauge, 
\begin{equation}
    D_{C,\mu\nu}(q) = \frac{\iu}{\vb{q}^2}v_\mu v_\nu +\frac{\iu}{q^2} \mathbb{P}_{\mu\nu}  ~, 
\end{equation}
where $\mathbb{P}_{\mu\nu}= -g^\perp_{\mu\nu}+q_{\perp \mu} q_{\perp \nu} /q_\perp^2$. 
We may compare this to Feynman gauge where one has instead 
\begin{equation}
    \begin{split}
    D_{F,\mu\nu}(q) &= (-1)\times \frac{\iu}{q^2} g_{\mu\nu} \\
                    &=(-1)\times \frac{\iu}{q^2} v_\mu v_\nu   -  \frac{\iu}{q^2} g^\perp_{\mu\nu} ~.
    \end{split}                  
\end{equation}
One can clearly focus on the terms proportional to $v_\mu v_\nu$ in Feynman gauge which must ({\it i}) produce the $q_{\perp \mu} q_{\perp \nu} /q_\perp^2$ terms in the projector $\mathbb{P}_{\mu\nu}$ and ({\it ii}) reproduce the longitudinal terms proportional to $v_\mu v_\nu$ in Coulomb gauge. 

For problem-({\it ii}) we first discuss the computation of the longitudinal modes in Coulomb gauge. One may readily check that the longitudinal contributions to diagram $\mathcal{A}_{a}^{(1)}$ vanishes by contour integration (diagrams $\mathcal{A}_{c}^{(1)}$ and $\mathcal{A}_{e}^{(1)}$ are trivially zero), because the longitudinal photon propagator has no poles in the complex-$\omega$ plane. The diagrams $\mathcal{A}_{b}^{(1)}$ and $\mathcal{A}_{d}^{(1)}$, by way of contrast, receive a contribution from the ``arc at infinity'' because there is only one double-line propagator leading to a $1/\omega$ fall-off as $\omega\rightarrow \infty$. The result is 
\begin{equation}
     \mathcal{A}_{b,C_L}^{(1)}= \iu e^2 \int \frac{\dd^4 q}{(2\pi)^4}\bigg[\frac{1}{\omega+\iu 0}\mel{B}{\mathcal{O}(\vb{k}) \frac{1}{-H_A+\iu 0}\mathbb{P}_{\slashed{A}}\rho(-\vb{q}) \rho(\vb{q})}{A}\bigg] \frac{1}{q^2} ~.
\end{equation}
A similar result holds for $\mathcal{A}_{d,C_L}^{(1)}$. 

To relate the Feynman and Coulomb gauge propagators it is convenient to make use of the identity 
\begin{equation}
    v_\mu= \frac{q_\mu - q_{\perp\mu}}{\omega+\iu 0}~,
\end{equation}
where the sign of the infinitesimal imaginary part is necessary to avoid pinches in the complex plane. Using $g_{\mu\nu} = v_{\mu}v_\nu + g^\perp_{\mu\nu}$ we may then write 
\begin{equation}
    g_{\mu\nu} = -\frac{q_\mu q_\nu}{(\omega+\iu 0)^2} + \frac{v_\mu q_\nu +v_\nu q_\mu}{(\omega+\iu 0)} +  \qty{g_{\mu\nu}^\perp + \frac{q_{\perp \mu} q_{\perp \nu} }{(\omega+\iu 0)^2}} ~.
\end{equation}
This representation is convenient because the term in curly braces reproduces the Coulomb-gauge result for transverse photons after performing the contour integral over $\dd \omega$, which amounts to the replacement $(\omega+\iu 0)^2\rightarrow - q_\perp^2$. This then solves problem-({\it i}) from above.

Let us turn our attention to problem-({\it ii}). We have already studied the propagator $q_\mu q_\nu /q^4$ when discussing $R_\xi$ gauge. The case of $q_\mu q_\nu /(q^2\times(\omega+\iu0)^2 )$ is slightly different because of the double-pole at $\omega=0$. In particular beneath 
\cref{Ab-xi-1} we dropped a term linear in $\omega$ from $\mathcal{A}^{(1)}_b$ noting that it would integrate to zero. This term now gives instead,  $\omega/(\omega+\iu 0)^2 = 1/(\omega+\iu 0) = {\rm PV}(1/\omega)+ (-\iu \pi) \delta(\omega)$ which does not integrate to zero when integrated against an even function. The result is that the terms involving $(\iu)\times q_{\mu} q_\nu /[q^2(\omega+\iu 0)^2]$ give
\begin{equation}
    \label{coulomb-contribution}
     (-1)\times \iu e^2 \int \frac{\dd^4 q}{(2\pi)^4}\bigg[\frac{1}{\omega+\iu 0}\mel{B}{\mathcal{O}(\vb{k}) \frac{1}{-H_A+\iu 0}\mathbb{P}_{\slashed{A}}\rho(-\vb{q}) \rho(\vb{q})}{A}\bigg] \frac{1}{q^2} = (-1)\times \mathcal{A}_{b,C_L}^{(1)}~,
\end{equation}
which is related to the longitudinal (Coulomb mode) contribution $\mathcal{A}_{b,C_L}^{(1)}$. Again, a similar result holds for $\mathcal{A}_{d,C_L}^{(1)}$.

In order to study the terms involving $(v_\mu q_\nu+ v_\nu q_\mu)$ let us define the ``mixed propagator''  
\begin{equation}
    D_{M,\mu\nu} = (-1)\times\frac{\iu}{q^2} \frac{1}{\omega+\iu0}\qty(v_\mu q_\nu+ v_\nu q_\mu)~.
\end{equation}
Consider first the contribution to $\mathcal{A}_{a}^{(1)}$ using $D_{M,\mu\nu}$ in place of $D_{\mu\nu}$ which we denote by $\mathcal{A}_{a,M}^{(1)}$, 
\begin{equation}
    \begin{split}
        \mathcal{A}_{a,M}^{(1)} =  -\iu e^2 \int \frac{\dd^4 q}{(2\pi)^4}&\bigg[\frac{1}{\omega+\iu 0}\mel{B}{\rho(-\vb{q})\frac{1}{\omega-H_B+\iu 0} \mathcal{O}(\vb{k}) \rho(\vb{q})}{A} \\
         &\hspace{0.15\linewidth}+ \frac{1}{\omega+\iu 0}\mel{B}{\rho(-\vb{q}) \mathcal{O}(\vb{k}) \frac{1}{\omega-H_A+\iu 0} \rho(\vb{q})}{A} \bigg]~.
    \end{split}
\end{equation}
It is useful to re-write this as, 
\begin{equation}
    \label{amplitude-a-general}
    \begin{split}
        \mathcal{A}_{a,M}^{(1)} =  -\iu e^2 \int \frac{\dd^4 q}{(2\pi)^4}& \frac{1}{q^2}\bigg[ \frac{1}{\omega+\iu 0}\mel{B}{\mathcal{O}(\vb{k})\rho(-\vb{q}) \frac{1}{\omega-H_A+\iu 0} \rho(\vb{q})}{A}\\
         &\hspace{0.125\linewidth}+ \frac{1}{\omega+\iu 0}\mel{B}{\rho(-\vb{q})\frac{1}{\omega-H_B+\iu 0}  \rho(\vb{q})\mathcal{O}(\vb{k})}{A} \bigg] \\
         -\iu  e^2 \int \frac{\dd^4 q}{(2\pi)^4}&\frac{1}{q^2}\bigg\{ \frac{1}{\omega+\iu 0}\mel{B}{\qty[\rho(-\vb{q}),\mathcal{O}(\vb{k})] \frac{1}{\omega-H_A+\iu 0} \rho(\vb{q})}{A} \\
         &\hspace{0.125\linewidth}+ \frac{1}{\omega+\iu 0}\mel{B}{\rho(-\vb{q})\frac{1}{\omega-H_B+\iu 0}  \qty[\mathcal{O}(\vb{k}),\rho(\vb{q})]}{A} \bigg\}~,
    \end{split}
\end{equation}
where all commutator dependence is between the curly braces. The terms in square brackets are organized such that the first line contains EM currents that act before the operator and the second line EM currents that act after the operator. Let us focus on the first line which combines with $\mathcal{A}_{b,M}^{(1)}$ and $Z_A^{(1)}$. The seagull terms in $\mathcal{A}_{c,M}^{(1)}$ give zero when contracted with $v_\mu$ and so we have for the rest of the terms (focusing on those involving interactions with $A$ before the external operator)
\begin{equation}
    \begin{split}
    \mathcal{A}^{(1)}_{b,M} + \frac12 Z^{(1)}_{A,M} \mathcal{A}^{(0)} =  -\iu e^2 \int \frac{\dd^4 q}{(2\pi)^4}&\frac{1}{q^2}\bigg[\frac{2}{\omega+\iu 0}\mel{B}{\mathcal{O}(\vb{k}) \frac{1}{-H_A+\iu 0}\mathbb{P}_{\slashed{A}}\rho(-\vb{q}) \rho(\vb{q})}{A}\\
    &\hspace{0.05\linewidth}- \frac{1}{\omega+\iu 0}\mel{B}{\mathcal{O}(\vb{k})\rho(-\vb{q}) \frac{1}{\omega-H_A+\iu 0} \rho(\vb{q})}{A}  \bigg] ~.
    \end{split}
\end{equation}
Notice that when combined with the first line in the square brackets of \cref{amplitude-a-general}  this gives $2 \times \mathcal{A}^{(1)}_{b,C_L}$. Finally, adding this to $(-1)\times \mathcal{A}^{(1)}_{b,C_L}$ from \cref{coulomb-contribution} we find the same result as would be obtained in Coulomb gauge for diagram-$(b)$. Performing the same analysis with the second line in the square brackets of \cref{amplitude-a-general} and $\mathcal{A}^{(1)}_{d,M} + \frac12 Z^{(1)}_{B,M} \mathcal{A}^{(0)}$ leads to (in the case where $[\rho,\mathcal{O}]=0$), 
\begin{equation}
    \mathcal{A}_{a,M}^{(1)}+\mathcal{A}^{(1)}_{b,M} + \frac12 Z^{(1)}_{A,M} \mathcal{A}^{(0)} + \mathcal{A}^{(1)}_{d,M} + \frac12 Z^{(1)}_{B,M} \mathcal{A}^{(0)}  = \mathcal{A}^{(1)}_{b,C_L} + \mathcal{A}^{(1)}_{d,C_L}~.
\end{equation}
This solves problem-({\it ii}) from above because $\mathcal{A}^{(1)}_{c,C}=\mathcal{A}^{(1)}_{e,C}=0$ and $\mathcal{A}^{(1)}_{a,C}= \mathcal{A}^{(1)}_{a,C_T}$ only involves transverse photons such that $\mathcal{M}^{(1)}_C=\mathcal{M}^{(1)}_F$ when summing over all of the Feynman diagrams. 

So far in our discussion we have neglected the commutators $[\rho(\vb{q}),\mathcal{O}(\vb{k})]$ in \cref{amplitude-a-general}. These terms are non-zero for charged operators\footnote{For operators involving derivatives there are further seagull terms at the weak vertex we have not included.} and are relevant for weak interaction phenomenology. Since charged currents mix the gauge properties of the hadronic and leptonic sectors, we must also study diagrams involving photon exchange with the charged lepton (or other charged hadron) i.e., the box diagrams discussed above. Doing so one finds that the contribution from $D_{M,\mu\nu}$ where $v_\mu$ contracts with the hadronic part of the diagram and $q_\nu$ with the leptonic part of the diagram subtracts off the commutator terms in \cref{amplitude-a-general}. To see this one requires the result that  $[\rho(\vb{q}),\mathcal{O}(\vb{k})]= \mathcal{Q}_\mathcal{O}\mathcal{O}(\vb{q}+\vb{k})$ and to use the on-shell identity $-\slashed{q} v(p_e)= (-\slashed{p}_e-\slashed{q}-m)v(p_e)$ for the leptonic part of the diagram which is related to diagrammatic proofs of the Ward identity. 

\section{Degenerate perturbation theory \label{app:deg-PT}}
It is a common occurrence when considering bound state spectra that closely spaced levels spoil the convergence of perturbation theory. The ``EPI scheme'' discussed in the main text corresponds to standard non-degenerate perturbation theory, however it is straightforwardly generalized to include a manifold of closely spaced energy levels using degenerate perturbation theory. 
The ``1PI scheme'' is then a particular choice in which all single particle states (including the entire bound-state spectrum) are included in the quasi-degenerate manifold. 
Here we briefly outline  how to extend the EPI scheme when the initial or final nucleus is part of an almost degenerate subspace, $S$, of states. Firstly we extend the definition of EPI to this whole space so that the full two point function $G^{(S)}_{ij }$  between any two states $i$ and $j$ in this subspace is. Removing some of the factors of $i$ we write

\begin{equation}
G^{(S)}_{ji} =D_{ji} +D_{jk'} \left[\Sigma_{k'k} \right] D_{ki} +... = \left[ {1 \over 1-D \Sigma}\right]_{jk}D_{ki}=D_{jk} \left[ {1 \over 1-\Sigma D} \right]_{kj} =\left[ 1 \over D^{-1}-\Sigma \right]_{ij}~,
\end{equation}
where repeated indices are summed over states in the almost degenerate subspace and
\begin{equation}
D_{ji}(p^0)={\delta_{ji}  \over p^0-E_i}~,
\end{equation}
is the free propagator. Of course $G$ and $\Sigma$ also depend on $p^0$. $\Sigma$ has a perturbative expansion in the electromagnetic fine structure constant $\alpha$ that begins at $O(\alpha)$.
For convenience we have removed a factor of $\iu$ from the conventional definitions (e.g. $D$ is $-\iu$ times the usual free propagator, $\Sigma$ is $\iu$ times the usual definition, and similarly for $G$). In the above $\Sigma $ is the subspace EPI  self energy matrix that takes into account electromagnetic corrections.   It has a perturbative series in the electromagnetic fine structure constant and the first term is linear in $\alpha$.  

Taking these electromagnetic corrections into account the corrected energy eigenvalues for states $E_i^{\rm phys.}$ in this subspace occur at values of $p^0$ that satisfy, ${\rm det}[ D(p^0)^{-1}-\Sigma(p^0) ]=0$.

In general the initial and final states ($A$ and $B$) will have their own degenerate subspaces $S_A$ and $S_B$.   For a transition between the states $A^{\rm phys.}$ in $S_A$ and $B^{\rm phys.}$ in $S_B$ we  need to find the physical states $A^{\rm phys.}$ and  $B^{\rm phys.}$ and which include electromagnetic corrections. Let's first focus first on the initial states and expand the denominator of $G^{(S_A)}$ about $p^0=E_A^{\rm phys.}$

\begin{equation}
D^{-1}-\Sigma=(p^0-E_A^{\rm phys.}+E_A^{\rm phys.}-E_A) -\Sigma(E_A^{\rm phys.})-(p^0-E_A^{\rm phys.})\Sigma'(E_A^{\rm phys.})+...
\end{equation}
The   matrix
\begin{equation}
 M=(E_A^{\rm phys.}-E_A)  -\Sigma^{(S_A)}(E_A^{\rm phys.})
\end{equation}
has one zero eigenvalue corresponding the physical initial state. Clearly we want to work with a basis of states for the subspace $S$ where $M$ is diagonal. So henceforth lets do so. For LSZ we will need.
\begin{equation}
{\rm lim}_{p^0 \rightarrow E_A^{\rm phys.}}  \left[ G^{(S_A)}_{k l} (p^0- E_A^{\rm phys.}) \right] ={\rm lim}_{p^0 \rightarrow E_A^{\rm phys.}}\left[ {1 \over 1+M / (p^0-E_A^{\rm phys.}) -\Sigma'(E_A^{\rm phys.})} \right]_{kl}
\end{equation}
Taking the limit it can be seen that this matrix is diagonal and the only non zero entry in the diagonal is the one corresponding to the physical state $A$. Explicitly

\begin{equation}
{\rm lim}_{p^0 \rightarrow E_A} \left[ G^{(S_A)}_{k l} (p^0- E_A^{\rm phys.}) \right] = {\delta_{k |A \rangle}  \delta_{j |A \rangle}  \over  \left(1-\Sigma'(E_A^{\rm phys.})_{|A \rangle |A \rangle} \right)}
\end{equation}
An identical set of equations holds for the final state and its almost degenerate subspace.  Since $\Sigma$  is order $\alpha$ if we work to that order we can replace $E_A^{\rm phys.}$ by $E_A$ in the argument of $\Sigma'$.

So the upshot is that the LSZ prescription is to rotate to the  basis where the matrices $M$  for the initial and final almost degenerate subspaces  are diagonal with a zero in the slot corresponding to the  initial   and final states and then the usual formulas apply with (at leading order in the electromagnetic coupling), $\sqrt{Z_A} \simeq 1+\Sigma'(E_A)_{|A \rangle |A \rangle} /2$ and $\sqrt{Z_B} \simeq 1+\Sigma'(E_B)_{|B \rangle |B \rangle} /2$. Note that in general the matrices $\Sigma'$ are different for the initial and final almost degenerate subspaces.

A similar situation occurs in degenerate perturbation theory however one must take into account that the energy shift vanishes at $O(e)$ in QED because if we write $|A)=\ket{A} + e |A)^{(1)} + e^2 |A)^{(2)} + \ldots$ then $\langle A| A)^{(1)}=0$ because $\ket{A}$ has the photonic vacuum whereas $|A)^{(1)}$ contains a $\ket{1\gamma}$ state. As a result we must work to second order to determine the appropriate energy shift for the states $|A)$ and $|B)$. Doing so, one finds that to avoid small denominators in perturbation theory, an energy-dependent (and therefore state-dependent) matrix must be diagonalized, just like $\Sigma(p_0= E_{A,i})$ described above.

\section{Nuclear structure correction \label{app:delta-NS}}
In this section we make contact with existing literature on the nuclear structure dependent correction, $\delta_{\rm NS}$, the Fermi function $F(Z,E)$, and the shape function $C(Z,E)$. These quantities, which appear in the theory of superallowed beta decays are important for a consistent analysis of radiative corrections.

Let us begin with the mapping to the notation of Ref.~\cite{Seng:2021syx}. It is easiest to make connection prior to any subtractions i.e., in terms of $V^{(1)}$ and $B^{(1)}$. Since we have included only the vector part of the current, the mapping is given by, 
\begin{align}
    V^{(1)} &\rightarrow  \delta \mathfrak{M}_3 +  \delta\mathfrak{M}_2 \\
    B^{(1)}         &\rightarrow  \delta\mathfrak{M}_{\gamma W}^a~. 
\end{align}
The term $\delta\mathfrak{M}_3=0$ in the isospin conserving limit (in the infinite target-mass limit). We have used ``$\rightarrow$'' instead of ``$=$'' due to some technical differences. For example, the corrections in Ref.~\cite{Seng:2021syx} are defined with a propagating $W$-boson, whereas we have worked in an effective theory at the level of nucleons. Another difference is that the outer corrections in Ref.~\cite{Seng:2021syx} are following the classic Sirlin conventions, whereas our $\tilde{O}_{\rm virt.}^{(1)}$ is defined in terms of a heavy particle effective field theory amplitude as in Refs.~\cite{Hill:2023acw,Cirigliano:2023fnz,Cirigliano:2024nfi}. Finally, note that we have studied the renormalization of $\mathcal{O}(\vb{k})= \mathcal{J}^-_0$, whereas Refs.~\cite{Sirlin:1977sv,Seng:2021syx} consider $\mathcal{J}_\mu^-$ which also includes the spatial component of the weak current. 

\vfill 
\pagebreak 

Let us further identify the Fermi function and shape factor in our result. Consider $\tilde{B}^{(1)}$, which we repeat here for convenience, 
\begin{equation}
    \begin{split}
    \tilde{B}^{(1)} = &\frac{1}{\bar{u}(p_\nu)\Gamma v(p_e)\mel{B}{\mathcal{O}(\vb{k})}{A}} \times e^2 \int \frac{\dd^4q}{(2\pi)^4 } D^{\mu\nu}(q) \bar{u}(p_\nu) \Gamma\frac{1}{-\slashed{p}_e-\slashed{q}-m} \gamma_\nu v(p_e)  \\
    & \hspace{0.0\linewidth}\times\bigg[ -(2\pi \iu) v_\mu \delta(\omega) \mel{B}{\rho(\vb{q})\mathcal{O}(\vb{k}-\vb{q})}{A} + \\
     & \hspace{0.05\linewidth} \mel{B}{\mathscr{J}_\mu^{(-)}(\vb{q}) \frac{1}{-\omega-H_B}\mathcal{O}(\vb{k}-\vb{q})}{A}  + \mel{B}{\mathcal{O}(\vb{k}-\vb{q}) \frac{1}{\omega-H_A}\mathscr{J}_\mu^{(+)}(\vb{q})}{A}\bigg ]~.
     \end{split}
\end{equation}
The first term with $\delta(\omega)$ is separately gauge invariant (in arbitrary non-covariant gauge), and remains gauge invariant for an arbitrary state inserted between $\rho(\vb{q})$ and $\mathcal{O}(\vb{k}-\vb{q})$. More precisely \cite{Plestid:2024ygd}, 
\begin{equation}
    D^{\mu\nu} (q) v_\nu \delta(\omega) = \frac{\iu}{\vb{q}^2} v^\mu \delta(\omega)  ~.   
\end{equation}
Let us insert the ground state, and focus only on this term
\begin{equation}
    \begin{split}
    e^2 &\int \frac{\dd^3q}{(2\pi)^3 }  \bar{u}(p_\nu) \Gamma\frac{1}{-\slashed{p}_e-\slashed{q}-m} \frac{\mel{B}{\rho(\vb{q})}{B}}{\vb{q}^2} \gamma_0 v(p_e) \mel{B}{\mathcal{O} (\vb{k}-\vb{q})}{A}  \\
    \rightarrow~ &\int \frac{\dd^3q}{(2\pi)^3 } \bar{u}(p_\nu) \Gamma ~\mathcal{V}_{p_e}(\vb{q})  \mel{B}{\mathcal{O} (\vb{k}-\vb{q})}{A} ~.
    \end{split}
\end{equation}
where we have identified the object on the first line as the $O(e^2)$ contribution to the outgoing positron wavefunction, $\mathcal{V}_{p_e}(\vb{q})$,  computed with in the field of nucleus-$B$,  convolved with the tree-level weak nuclear matrix element  $\mel{B}{\mathcal{O} (\vb{k}-\vb{q})}{A}$.  This matches the conventional definition of $F(Z,E)\times C(Z,E)$. 

We may then identify $\tilde{B}^{(1)}+ \tilde{V}^{(1)}$ with the $O(\alpha)$ contributions to $F(Z,E)\times C(Z,E) \times(1+\delta_{\rm NS})$. This quantity includes isospin breaking, however following the conventions of the literature, this one-loop isospin breaking correction is folded into $\delta_{\rm NS}$ which is then a gauge-invariant correction. Let us define, 
\begin{equation}
    \begin{split}
    \tilde{B}^{(1)}_{{\slashed{B}}} = &\frac{1}{\bar{u}(p_\nu)\Gamma v(p_e)\mel{B}{\mathcal{O}(\vb{k})}{A}} \times e^2 \int \frac{\dd^4q}{(2\pi)^4 } D^{\mu\nu}(q) \bar{u}(p_\nu) \Gamma\frac{1}{-\slashed{p}_e-\slashed{q}-m} \gamma_\nu v(p_e)  \\
    & \hspace{0.0\linewidth}\times\bigg[ -(2\pi \iu) v_\mu \delta(\omega) \mel{B}{\rho(\vb{q})\mathbb{P}_{\slashed{B}}\mathcal{O}(\vb{k}-\vb{q})}{A} + \\
     & \hspace{0.05\linewidth} \mel{B}{\mathscr{J}_\mu^{(-)}(\vb{q}) \frac{1}{-\omega-H_B}\mathcal{O}(\vb{k}-\vb{q})}{A}  + \mel{B}{\mathcal{O}(\vb{k}-\vb{q}) \frac{1}{\omega-H_A}\mathscr{J}_\mu^{(+)}(\vb{q})}{A}\bigg ]~,
     \end{split}
\end{equation}
then $\tilde{V}^{(1)}+ \tilde{B}^{(1)}_{{\slashed{B}}}$ can be mapped onto $\delta_{\rm NS}$.

\section{Ultraviolet sensitivity and dynamic range \label{app:dynamic-range}}
In the main text we have developed a formulation for computing objects like $\Sigma_2'$ of a bound-state. The expressions we encounter contain integrals which formally span from $|\vb{q}|=0$ up to $|\vb{q}|\sim \infty$. Even in renormalizable theories, like the example we discuss below, these integrals can receive large contributions from regions of integration outside the domain of validity of the effective description used to model the bound state; one may say that a numerical calculation of the matrix elements would require large ``dynamic range''. Since a large dynamic range is numerically challenging, it is desirable to remove the sensitivity to regions of large momentum transfer in the integral, and we outline a strategy to achieve this goal below. 

For concreteness in what follows, we will discuss the case of non-relativistic bound states relevant for atoms, nuclei, and heavy-heavy mesons such as the $\Upsilon$ or $J/\psi$. In order for a non-relativistic bound state to form it is crucial that the kinetic energy operator $-\nabla^2/2M$ is included in the zeroth order Hamiltonian. For definiteness let us write the Lagrangian of a nucleus as\footnote{In principle one could include dynamical pions instead of a non-local potential, but we use the following model for illustration purposed and view this detail as an unnecessary complication.}
\begin{equation}
    \mathcal{L}_S(x) =  \psi^\dagger(x)\qty( \iu D_t  -\frac{\vb{D}^2}{2M} )\psi(x)  + \frac12 \int \dd^3 y\qty[\rho(\vb{x},t) V(|\vb{x}-\vb{y}|) \rho(\vb{y},t)]
\end{equation}
where $i\in \{n,p\}$ the covariant derivatives are defined by 
\begin{align}
    D_t &= \partial_t + \iu e Q_i A_0~,\\
    \vb{D} &= \vb*{\nabla}+ \iu e Q_i \vb{A}~, 
\end{align}
with $Q_i=0$ for a neutron, and $Q_i=1$ for a proton. 
This leads to the density and current operators,\!\footnote{We define our operators with momentum flowing into the bound state, which differs from a conventional definition in terms of the Fourier transform of $\rho(\vb{x})$ or $\vb{J}(\vb{x})$. } with $a_{\vb{p}}$ the annihilation operator for a proton with momentum $\vb{p}$, 
\begin{align}
    \rho(\vb{q}) &= \int \frac{\dd^3p }{(2\pi)^3}  a^\dagger_{\vb{p}+\vb{q}}a_{\vb{p}} ~,\\
    \vb{J}(\vb{q}) &=\int \frac{\dd^3p }{(2\pi)^3}  \frac{2\vb{p}+\vb{q}}{2M} a^\dagger_{\vb{p}+\vb{q}}a_{\vb{p}}~,
\end{align}
and a Hamiltonian, 
\begin{equation}
    \label{H-S-NR}
    H_S = \int \dd^3x  ~\psi^\dagger(\vb{x})\qty[-\frac{\vb*{\nabla}^2}{2M}]\psi(\vb{x})+ \frac12 \int \dd^3 y\qty[\rho(\vb{x}) V(|\vb{x}-\vb{y}|) \rho(\vb{y})]~,
\end{equation}
where all quantities are evaluated at $t=0$. 

Let us consider $\Sigma_2'$ of nucleus $A$ working in Coulomb gauge. In this gauge only transverse photons contribute and we encounter the integral (after performing the $\omega$ integral by closing the contour in the upper half plane), 
\begin{equation}
    \Sigma_2 = e^2 \int \frac{\dd^3q}{(2\pi)^3}  \frac{1}{2|\vb{q}|} \mel{A}{J_i(\vb{-q}) \qty(\frac{1}{|\vb{q}| + \hat{H}-\iu 0})^2 J_j(\vb{q})}{A} \mathbb{P}_{ij}~,
\end{equation}
where $\mathbb{P}_{ij}= \delta_{ij}- q_i q_j/\vb{q}^2$. Using the explicit form for $\vb{J}(\vb{q})$ and the fact that $q_i \mathbb{P}_{ij}=0$, we arrive at, 
\begin{equation}
    \label{Sigma-2-coulomb-gauge}
    \Sigma_2 = e^2 \int \frac{\dd^3q}{(2\pi)^3} \frac{\dd^3p}{(2\pi)^3}  \frac{\dd^3k}{(2\pi)^3}  \frac{1}{2|\vb{q}|} \mel{A}{a^\dagger_{\vb{k}-\vb{q}} a_{\vb{k}}\qty(\frac{1}{|\vb{q}| + \hat{H}-\iu 0})^2 a^\dagger_{\vb{p}+\vb{q}} a_{\vb{p}}}{A} \mathbb{P}_{ij} \frac{p_i k_j}{M^2}~,
\end{equation}
As $|\vb{q}|\rightarrow \infty$ this integral is dominated by intermediate states in which a single nucleon (since $\vb{J}(\vb{q})$ is a one-body operator) is knocked out and carries momentum $\vb{p}+\vb{q}$ i.e., by the quasi-elastic peak. Neglecting the small excitation energy of the residual nuclear system we have  $H\ket{N(\vb{p}+\vb{q});A-1} \simeq (2M)^{-1}(\vb{p}+\vb{q})^2\ket{N(\vb{p}+\vb{q});A-1}$. The integral in \cref{Sigma-2-coulomb-gauge} is therefore convergent in the ultraviolet, but is sensitive to scales of $|\vb{q}|\sim 2 M$ which are much higher than the cut-off (or regime of validity) of the non-relativistic description which underlies \cref{H-S-NR}. 

This is a well known property of non-relativistic field theories (such as NRQCD or NRQED) \cite{Luke:1996hj,Manohar:1997qy}. Since we only expect approximations to nuclear matrix element to be reliable for $|\vb{q}|\lesssim 500~{\rm MeV}$ it is desirable to remove the sensitivity to $|\vb{q}|\sim M$. This can be achieved by subtracting the contribution from one-nucleon knockout,
\begin{equation}
    \qty[\Sigma_2' ]_{\rm subtr.}= e^2 \int \frac{\dd^3q}{(2\pi)^3} \frac{\dd^3p}{(2\pi)^3}  \frac{\dd^3k}{(2\pi)^3}  \frac{1}{2|\vb{q}|} \mel{A}{ O_S(\vb{k},\vb{p},\vb{q}) }{A} \mathbb{P}_{ij} \frac{p_i k_j}{M^2} 
\end{equation}
where the subtracted operator is 
\begin{equation}
    \begin{split}
    O_S(\vb{k},\vb{p},\vb{q})  =&a^\dagger_{\vb{k}} a_{\vb{k}+\vb{q}}\qty(\frac{1}{|\vb{q}| + \hat{H}-\iu 0})^2 a^\dagger_{\vb{p}+\vb{q}} a_{\vb{p}}\\
    &\hspace{0.1\linewidth}-a^\dagger_{\vb{k}}  \qty(\frac{1}{|\vb{q}| + (\vb{q}+\vb{p})^2/2M-\iu 0})^2 a_{\vb{p}}~(2\pi)^3\delta^{(3)}(\vb{p}-\vb{k}). 
    \end{split}
\end{equation}

We must add back what we have subtracted, however this piece factorizes into an analytically calculable piece, and simple matrix element inside the nucleus, 
\begin{equation}
    \Sigma_2'= \qty[\Sigma_2' ]_{\rm subtr.} + \frac23 e^2 \langle \vb{v}_p^2 \rangle_A  
    \int \frac{\dd^3q}{(2\pi)^3} \frac{1}{2|\vb{q}|} \qty(\frac{1}{|\vb{q}| + \vb{q}^2/2M-\iu 0})^2  + O(\vb{v}_p^4)~,
\end{equation}
where we have kept only the leading-in-$\vb{v}_p$ contribution, and where $\langle \vb{v}_p^2 \rangle = \langle \sum_i \vb{v}_i^2 \rangle$ where the sum runs over all protons in the nucleus. 
Now $[\Sigma_2']_{\rm subtr.}$ is sensitive only to nuclear scales, and the extra integral can be calculated analytically. Since $|\vb{p}|/M$ is always small, one can expand the integral around $|\vb{p}|=0$. 

Not all of these ``high momentum" contributions start at order ${\bf p}^2$. For example performing a similar analysis for $\tilde{V}_F^{(1)}-\tilde{V}_C^{(1)}$ gives
\begin{equation}
\left[\tilde{V}_F^{(1)}-\tilde{V}_C^{(1)}\right]_{\rm subtr}={e^2 \over 2M}\int{d^3q \over (2 \pi)^3}{1 \over 2 |{\bf q|}}\left({1 \over |{\bf q}|+{\bf q}^2/2M +\iu 0}\right)+O( v_{\bf p}^2)~.
\end{equation}
More generally one may study non-relativistic effective theories that yield divergent integrals and demand renormalization. These ultraviolet divergences are qualitatively similar to the issues discussed above. It is clear that one may perform appropriate subtractions yielding many-body matrix elements that are convergent in the ultraviolet. All renormalization (such as mass renormalization of the proton) can then be handled analytically, e.g. in dimensional regularization. It is important to emphasize that any and all renormalization can be performed in the single-body sector\footnote{If considering multi-body operators that contribute to the current then the single-body sector generalizes to the few body-sector.} at the level of the constituent particles, and that all other shifts to bound-state masses will be finite and calculable predictions of the theory.

\end{fmffile}

\bibliographystyle{apsrev4-1}
\bibliography{condensed}

\end{document}